\documentclass[prd,showpacs,preprintnumbers]{revtex4-1}
\usepackage{feynmp}
\usepackage{graphicx}
\usepackage{hyperref}
\usepackage{amsfonts}
\usepackage{amsmath,amssymb}
\usepackage{feynmp}
\usepackage{empheq}
\usepackage{tikz}
\usepackage{pgfplots}
\usepackage{subfig}
\usepackage{caption}
\usetikzlibrary{calc,positioning,arrows,fit,intersections,decorations.pathmorphing,decorations.markings}

\usepackage{graphicx}
\usepackage{hyperref}
\usepackage{amsfonts}
\usepackage{amsmath,amssymb}
\usepackage{feynmp}
\usepackage{empheq}
\usepackage{tikz}
\usepackage{pgfplots}
\usepackage{subfig}
\usepackage{caption}
\usetikzlibrary{calc,positioning,arrows,fit,intersections,decorations.pathmorphing,decorations.markings}

 \def\be{\begin{equation}}
 \def\ee{\end{equation}}
 \def\bes{\begin{eqnarray}}
 \def\ees{\end{eqnarray}}

 \def\p{\partial}

 \def\2{\frac{1}{2}}
 \def\4{\frac{1}{4}}

 \def\nextl{ \nonumber\\ & &}

\catcode`\@=11
\def\@citex[#1]#2{%
\if@filesw \immediate \write \@auxout {\string \citation {#2}}\fi
\@tempcntb\m@ne \let\@h@ld\relax \def\@citea{}%
\@cite{%
  \@for \@citeb:=#2\do {%
    \@ifundefined {b@\@citeb}%
      {\@h@ld\@citea\@tempcntb\m@ne{\bf ?}%
      \@warning {Citation `\@citeb ' on page \thepage \space
undefined}}%
      {\@tempcnta\@tempcntb \advance\@tempcnta\@ne%
      \@tempcntb\number\csname b@\@citeb \endcsname \relax%
      \ifnum\@tempcnta=\@tempcntb 
it
        \ifx\@h@ld\relax%
          \edef \@h@ld{\@citea\csname b@\@citeb\endcsname}%
        \else%
          \edef\@h@ld{\ifmmode{-}\else--\fi\csname
b@\@citeb\endcsname}%
        \fi%
      \else
        \@h@ld\@citea\csname b@\@citeb \endcsname%
        \let\@h@ld\relax%
      \fi}%
    \def\@citea{,\penalty\@highpenalty\,}%
  }\@h@ld
}{#1}}

\def\@citeb#1#2{{[#1]\if@tempswa , #2\fi}}
\def\@citeu#1#2{{$^{#1}$\if@tempswa , #2\fi }}
\def\@citep#1#2{{#1\if@tempswa , #2\fi}}

\begin{document}
\title{Tree-level Correlators of scalar and vector fields in AdS/CFT}

\author{Savan Kharel}
 \email{skharel@tennessee.edu}
\author{George Siopsis}
 \email{siopsis@tennessee.edu}
\affiliation{%
Department of Physics and Astronomy,
The University of Tennessee,
Knoxville, TN 37996 - 1200, USA}
\date{July 2013}
\begin{abstract}
Extending earlier results by Paulos, we discuss further the use of the embedding formalism and Mellin transform in the calculation of tree-level correlators of scalar and vector fields in AdS/CFT. We present an iterative procedure that builds amplitudes by sewing together lower-point off-shell diagrams. Both scalar and vector correlators are shown to be given in terms of Mellin amplitudes. We apply the procedure to the explicit calculation of three-, four- and five-point correlators.
\end{abstract}

\maketitle

\section{Introduction}The Anti-de Sitter / Conformal Field Theory correspondence (AdS/CFT) has revealed important connections between 
 quantum gravity and gauge theory \cite{Maldacena:1997re,Witten:1998qj}. Even though AdS/CFT 
provides a prescription for the holographic computation of correlation functions in a strongly coupled gauge theory with a gravity dual, in practice, 
computing these correlation functions is quite difficult.
The conformal correlators are related to scattering amplitudes in AdS space. The latter are not defined in a standard fashion, as in Minkowski space, because AdS space does not admit asymptotic states which are needed for the standard definition of the $S$-matrix. Nevertheless, creation and annihilation operators can be defined in AdS space by changing the boundary conditions in the conformal boundary \cite{Penedones:2010ue}. The resulting scattering amplitudes in AdS space are then related to CFT correlation functions.

AdS scattering amplitudes are derived from Witten diagrams which are difficult to calculate in coordinate space \cite{Freedman:1998bj,Liu:1998ty,Freedman:1998tz,D'Hoker:1999ni,D'Hoker:1999pj}. There have been some interesting developments in the computation of such diagrams in momentum space \cite{Raju:2010by,Raju:2012zr,
Raju:2012zs,Raju:2011mp}. Working in momentum space entails taking a Fourier transform of the amplitude, which is well-suited for flat Minkowski space, but does not appear to be advantageous in AdS space. Another approach using Mellin representation of conformal correlation functions was  proposed in \cite{Mack:2009mi, Mack:2009gy, Penedones:2010ue}. In more recent work \cite{Paulos:2011ie, Fitzpatrick:2011ia}, it was shown that CFT correlation functions factorize on poles in a Mellin representation, which suggests that Witten diagrams can be computed via a set of Feynman rules, as is the case with correlation functions of field theories on Minkowski space in the momentum (Fourier) representation.

In the case of scalar fields, by taking the Mellin transform, one trades coordinates for Mandelstam invariants of the scattering amplitude. This does not extend straightforwardly to vector or general tensor fields because of the index structure. After taking a Mellin transform, one is still left with expressions which involve coordinates, as well as Mandestam invariants. The index structure complicates calculations which involve integrals over coordinates in AdS space. Our aim is to extend the results of \cite{Paulos:2011ie} and provide a general procedure for the calculation of Witten diagrams involving fields of arbitrary spin. We shall show that diagrams of vector fields can be written in terms of the same Mellin functions as scalar field diagrams. Our method is an iterative procedure that calculates a diagram of a certain order by sewing together lower-order diagrams. The index structure is dealt with by taking advantage of the conformal properties of the correlation functions. Extending our method to higher-spin fields is straightforward and will be reported on in the near future.

The outline of our discussion is the following. In section \ref{secI}, we review the basic ingredients in the embedding formalism which seems to be the most natural framework for Mellin
representation. In sections \ref{secII}, \ref{secIII}, and \ref{secIV}, we calculate explicitly three-, four-, and five-point amplitudes, respectively, for scalar fields with a cubic interaction as well as vector fields.
In section \ref{secV}, we discuss the calculation of a general $N$-point diagram from lower-order constituents.
Finally, in section \ref{secVI}, we summarize our conclusions.
Appendix \ref{Xintegral} contains all necessary integrals over AdS space together with their derivation.

\section{Basics}
\label{secI}

In this section, we review the basic ingredients to be used in our discussion. We adopt the notation used in \cite{Paulos:2011ie}, where further discussion can be found.

It is natural to use the embedding space formalism, which goes back to Dirac \cite{Dirac:1936fq} (also see \cite{Weinberg:2010fx}), as it provides  a convenient framework for the computation of Witten diagrams. The embedding is a $(d+2)$-dimensional space flat Minkowski psace ($\mathbb{M}_{d+2}$) with metric given by
\be
\label{eqn1}
ds^2 = dX_AdX^A = - (dX^0)^2 + (dX^1)^2 + \dots + (dX^d)^2 + (dX^{d+1})^2~.
\ee
The Euclidean AdS$_{d+1}$ space is defined as the hyperboloid $X^2=-R^2$,
where  $X^0>0$ , $X^A \in \mathbb{M}^{d+2}$.  Henceforth, we set $R=1$.

In this formalism, it is convenient to think of the conformal boundary of AdS as the space of null rays $P^A$ (with
$P^2=0$, and $P\sim \lambda P$). Then a correlation function of the dual CFT of weight $\Delta$ scales as $\mathcal{F}_\Delta (\lambda P) = \lambda^{-\Delta} \mathcal{F}_\Delta (P)$.
We will be interested in $n$-point correlation functions of the form $\mathcal{F} (P_1, P_2, \dots , P_n)$, and frequently use the notation
\be
\label{eqn3}
P_{ij}=-2P_i \cdot P_j \ .
\ee
We will use $X^A$, $Y^A$, etc., for points in the bulk, and $P^A$, $Q^A$, etc., for points on the boundary of AdS space.

The bulk to boundary propagator for a scalar field is given by
\be
\label{eqn4}
E (X,P)=\frac{\Gamma(\Delta)}{2 \pi^{d/2} \Gamma(1+\Delta-d/2) }
(-2 P \cdot X )^{-\Delta} ~.
\ee
The bulk to bulk propagator for a scalar field can be written as an integral over the boundary point $Q$ \cite{Penedones:2010ue}, 
\be 
\label{eqn6}
G(X,Y) = \int_{-i\infty}^{+i\infty} \frac{dc}{2\pi i} f_{\delta,0} (c)  \Gamma (d/2+c) \Gamma (d/2-c)\int_{\partial AdS} dQ  (-2X\cdot Q)^{-d/2-c} (-2Y\cdot Q)^{-d/2+c}~, \ee
where
\be
\label{eqn7}
f_{\delta,0} (c) = \frac{c \sin \pi c}{2\pi^{d+1} [ (\delta -d/2 )^2 -c^2] }\ . \ee
These expressions for the propagators are crucial for the factorization of amplitudes into lower-point diagrams. We are interested in calculating the $N$-point scalar amplitude
\be A^{(N, s)}= \langle \mathcal{O}_{\Delta_1}(P_1) \mathcal{O}_{\Delta_2}(P_2)  \cdots \mathcal{O}_{\Delta_N}(P_N) \rangle~, \ee
where $\mathcal{O}_\Delta$  is a conformal operator of scaling dimension $\Delta$.

Similarly, the bulk to boundary propagator for a vector field can be written as \cite{Paulos:2011ie},
\be
E_{MA} (X,P)  = D_{MA} (\Delta, P) E(X,P) = \frac{\Gamma(\Delta)}{2 \pi^{d/2} \Gamma(1+\Delta-d/2) } J_{MA}(-2 P \cdot X )^{-\Delta} 
\ee
where,
\be D_{MA} (\Delta , P) = \frac{\Delta -1}{\Delta} \eta_{M A} + \frac{1}{\Delta} \frac{\partial}{\partial P^{M}} P_{A} \ee
and $J_{MA} = \eta_{MA} - \frac{P_AX_M}{P\cdot X}$
has the property, $P^{M} J_{M A}= J_{M A} X^{A}=0$. 
$D_{MA}$ is an extremely convenient operator as it organizes and simplifies the index structure of vector amplitudes, allowing us to relate them to amplitudes of scalar fields.
In this regard, a useful identity is
\be\label{eq42i4} D_{MA} (\Delta , P)\frac{\partial}{\partial P_{A}} \mathcal{F}_{\Delta-1} (P) = 0~, \ee
where $\mathcal{F}_{\Delta-1}$ is a function of weight $\Delta -1$,  i.e., $\mathcal{F}_{\Delta -1} ( P) = \lambda^{-(\Delta -1)} \mathcal{F}_{\Delta -1}(P)$, and therefore $P\cdot \frac{\partial}{\partial P} \mathcal{F}_{\Delta -1} = - (\Delta -1) \mathcal{F}_{\Delta -1}$.

The bulk to bulk propagator for a vector field can be written as an integral over the boundary point $Q$, as in the scalar case,
\bes
\label{bbp}
G_{AB} (X,Y) &=&   \int_{-i\infty}^{+i\infty} \frac{dc}{2\pi i} f_{\delta,1} (c)  \Gamma (d/2+c) \Gamma (d/2-c) \nextl
\times \int_{\partial AdS} dQ\eta^{MN}D_{M A} (d/2+c, Q)D_{N B} (d/2-c,Q)  (-2X\cdot Q)^{-d/2-c} (-2Y\cdot Q)^{-d/2+c}~, \ees
where,
\be f_{\delta,1} (c) = f_{\delta,0} (c) \frac{\frac{d^2}{4}-c^2}{ (\delta - \frac{d}{2} )^2 -c^2}~. \ee
We are interested in calculating the $N$-point vector amplitude
\be A^{(N, v)M_1 \dots M_N, a_1\dots a_N}= \langle \mathcal{J}^{M_1,a_1}(P_1) \mathcal{J}^{M_2,a_2}(P_2)  \cdots \mathcal{J}^{M_N,a_N} (P_N) \rangle~, \ee
where $a_i$ ($i=1,\dots, N$) are gauge group indices. It should be pointed out that all current operators have dimension $\Delta = d-1$. However, we need to calculate off-shell amplitudes as well, because we are interested in sewing diagrams together in order to form higher-point amplitudes. The two legs to be sewn must be off shell, and have dimensions $\frac{d}{2} \pm c$, on account of \eqref{bbp}. Therefore, we will be generally working with arbitrary dimensions of the external legs of a $N$-point vector amplitude.

The Mellin transform of the above $N$-point amplitudes will be given in terms of Mandelstam invariants $\delta_{ij}$ ($i,j = 1,\dots, N$). They are defined with the properties
\be\label{eqMand} \delta_{ii} = 0 \ \ , \ \ \ \ \delta_{ij} = \delta_{ji} \ \ , \ \ \ \ \sum_{j=1}^N \delta_{ij} = \Delta_i ~.\ee

\section{Three-point Amplitudes}
\label{secII}

Having introduced all necessary ingredients, we now proceed to the explicit calculation of amplitudes, starting with the simplest amplitude.

\subsection{Scalar amplitudes}

The three-point amplitude for scalar fields of scaling dimensions $\Delta_i$ interacting via a cubic interaction of coupling constant $g$ is (Fig.\ \ref{3pt})
\be\label{eqn8} A^{(3,s)}(\Delta_1,P_1;\Delta_2,P_2;\Delta_3,P_3) =\frac{g}{\prod_i 2\pi^h \Gamma (\Delta_i +1 -d/2)} \mathcal{A}_3 (\Delta_1, P_1; \Delta_2, P_2; \Delta_3, P_3) \ee
where
\be \mathcal{A}_3 ( \{\Delta_i , P_i\}) \equiv \int_{ AdS} dX \prod_{i=1}^3 \Gamma (\Delta_i) (-2P_i\cdot X)^{-\Delta_i}~.\ee
\tikzset{
particle/.style={thick,draw=black},
gluon/.style={decorate, draw=black,
    decoration={coil,aspect=0}}
 }
\begin{center}
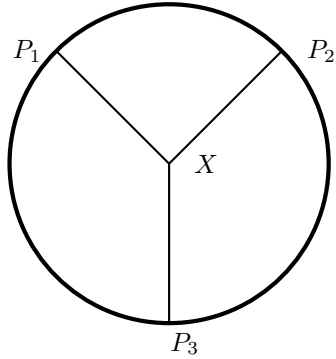

\begin{tikzpicture}[node distance=1.5cm and 1.5cm]
\coordinate[label={[xshift=-2pt]left:$P_1$}] (e1);
\coordinate[below right=of e1, ,label={[xshift=6pt]right:$X$}] (aux1);
\coordinate[above right=of aux1,label={[xshift=6pt]right:$P_2$}] (e2);
\coordinate[below=1/cos(45)*1.5cm of aux1,label={[xshift=6pt]below:$P_3$}] (aux2);

\draw[particle] (e1) -- (aux1);
\draw[particle] (aux1) -- (e2);

\draw[particle] (aux1) --  (aux2);
\draw  [ultra thick] (aux1) circle [radius=sqrt(1.5cm^2+1.5cm^2)];
\end{tikzpicture}
\captionof{figure}{The three-point scalar amplitude \eqref{eqn8}.}
\label{3pt}
\end{center}
The integral over the bulk vector $X^A$ is of the form \eqref{eqA12}. Using \eqref{eqA16}, we obtain
\be\label{eq1f} \mathcal{A}_{3}(\Delta_1,P_1;\Delta_2,P_2;\Delta_3,P_3) = \frac{\pi^{d/2}}{2} \mathcal{M}_3  \Gamma( \delta_{12})\Gamma (\delta_{23}) \Gamma (\delta_{13}) ( P_{12})^{-\delta_{12}} (P_{23})^{-\delta_{23}} (P_{13})^{-\delta_{13}} \ee
where
\be \mathcal{M}_3 = \Gamma \left( \frac{\Delta_{1}+\Delta_{2}+\Delta_{3} -d}{2}  \right) \ee
is the Mellin transform of the scalar three-point amplitude.
There are no remaining integrals, because the constraints \eqref{eqA11} completely fix the integration variables,
\be \delta_{12} = \frac{\Delta_1+\Delta_2-\Delta_3}{2} \ , \ \
\delta_{23} = \frac{\Delta_2+\Delta_3-\Delta_1}{2} \ , \ \
\delta_{31} = \frac{\Delta_1+\Delta_3-\Delta_2}{2}
\ee
which are the Mandelstam invariants \eqref{eqMan} for a three-point amplitude.

The three-point scalar amplitude is
\be\label{eq1fa} A^{(3,s)}(\Delta_1,P_1;\Delta_2,P_2;\Delta_3,P_3) = \frac{g\pi^{d/2}}{2 \prod_i 2\pi^{d/2} \Gamma (\Delta_i +1 -d/2)} \mathcal{M}_3   \prod_{i<j} \Gamma( \delta_{ij}) P_{ij}^{-\delta_{ij}}~. \ee

\subsection{Vector amplitudes}

Similarly, a three-point vector amplitude is given by (fig.\ \ref{3ptv2})
\be\label{eq20} A^{(3,v)M_1M_2M_3,a_1a_2a_3} (\Delta_1,P_1;\Delta_2,P_2;\Delta_3,P_3) = f^{a_1a_2a_3}\prod_{i=1}^3D^{M_iA_i} (\Delta_i,P_i) \mathcal{A}_{A_1A_2A_3} \ee
where,
\be\label{eq21} \mathcal{A}_{A_1A_2A_3} = \int_{AdS} dX I_{A_1A_2A_3} \prod_{i=1}^3 (-2P_i\cdot X)^{-\Delta_i}~, \ee
and the index structure is similar to a gauge theory three point vertex in flat space,
\be I_{A_1A_2A_3} = \eta_{A_1A_2} \left( K_{1A_3} - K_{2A_3} \right) + \cdots + \cdots~, \ee
with
\be K_A = (-2P\cdot X)^{\Delta} \frac{\partial}{\partial X^A} (-2P\cdot X)^{-\Delta}=- \Delta \frac{P_A}{P\cdot X} \ee
\tikzset{
particle/.style={decorate, draw=black,
    decoration={coil,aspect=0.08,segment length=3pt,amplitude=3pt}}}
\begin{center}
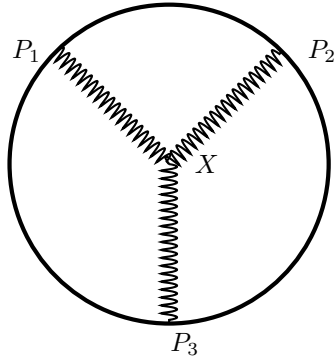

\begin{tikzpicture}[thick, node distance=1.5cm and 1.5cm]
\coordinate[label={[xshift=-2pt]left:$P_1$}] (e1);
\coordinate[below right=of e1, ,label={[xshift=6pt]right:$X$}] (aux1);
\coordinate[above right=of aux1,label={[xshift=6pt]right:$P_2$}] (e2);
\coordinate[below=1/cos(45)*1.5cm of aux1,label={[xshift=6pt]below:$P_3$}] (aux2);

\draw[particle] (e1) -- (aux1);
\draw[particle] (aux1) -- (e2);

\draw[particle] (aux1) --  (aux2);
\draw  [ultra thick] (aux1) circle [radius=sqrt(1.5cm^2+1.5cm^2)];
\end{tikzpicture}
\captionof{figure}{The three-point vector amplitude \eqref{eq20}.}
\label{3ptv2}
\end{center}
Thus, the three-point vector amplitude \eqref{eq21} is, explicitly,
\be\label{eq37a} \mathcal{A}_{A_1A_2A_3} = -\int_{AdS} dX \left[ \eta_{A_1A_2} \Delta_1\frac{P_{1A_3}}{P_1\cdot X} + \eta_{A_2A_3} \left( \Delta_2\frac{P_{2A_1}}{P_2\cdot X} - \Delta_3\frac{P_{3A_1}}{P_3\cdot X} \right) - (1\leftrightarrow 2) \right] \prod_{i=1}^3 \Gamma (\Delta_i)(-2P_i\cdot X)^{-\Delta_i} ~. \ee
As in the scalar case, the integral over the bulk vector $X^A$ is of the form \eqref{eqA12}. Using \eqref{eqA16}, we obtain
\bes
\mathcal{A}_{A_1A_2A_3} &=& \eta_{A_1A_2} P_{1A_3} \mathcal{A}_{3} (\Delta_1+1,P_1;\Delta_2,P_2; \Delta_3, P_3) \nonumber\\
&& + \eta_{A_2A_3} \left[ P_{2A_1} \mathcal{A}_{3} (\Delta_1,P_1;\Delta_2+1,P_2; \Delta_3, P_3) - P_{3A_1} \mathcal{A}_{3} (\Delta_1, P_1; \Delta_2, P_2; \Delta_3+1, P_3) \right] \nonumber\\
&& - (1\leftrightarrow 2) \ees
Thus, the vector amplitude is written in terms of scalar amplitudes.

As in the scalar case (Eq.\ \eqref{eq1f}), we may write this in terms of a Mellin amplitude,
\be\label{eq1fv} \mathcal{A}_{A_1A_2A_3} = \frac{\pi^{d/2}}{2}\mathcal{M}_{A_1A_2A_3}  \prod_{i<j} \Gamma\left( \delta_{ij} +\frac{1}{2} \right) P_{ij}^{-\delta_{ij}+\frac{1}{2}} \ee
where
\be \mathcal{M}_{A_1A_2A_3} = \Gamma \left( \frac{\Delta_{1}+\Delta_{2}+\Delta_{3} -d+1}{2}  \right) \left[ \mathcal{I} (1,2,3) + \mathcal{I} (2,3,1) + \mathcal{I} (3,1,2) \right] \ee
and
\be  \mathcal{I} (1,2,3) =
\frac{\eta_{A_1A_2}}{P_{12}} \left( \frac{1}{ \delta_{23} -\frac{1}{2}} \frac{P_{1A_3} }{P_{13}}  - \frac{1}{\delta_{13} -\frac{1}{2}} \frac{P_{2A_3}}{P_{23}} \right)  \ee
The above expressions are simplified if all legs are on shell. Setting $\Delta_1=\Delta_2=\Delta_3 = d-1$, we obtain
\be \mathcal{A}^{\mathrm{(on~shell)}}_{A_1A_2A_3} = \pi^{d/2}\Gamma \left( d-2  \right) \left[ \eta_{A_1A_2} \left( P_{23}P_{1A_3}  - P_{13} P_{2A_3} \right) + \cdots + \cdots \right] \prod_{j<j} \Gamma (d/2) P_{ij}^{-d/2}~.\ee
In order to use this amplitude in a higher-point diagram, it is convenient to eliminate terms that contain the coordinate that corresponds to the leg which is to be sewn (off-shell) with a free index (i.e., not in a dot product). Without loss of generality we choose the last leg, a practice that we will follow throughout.

Thus, we wish to eliminate terms containing $P_3^{A}$. To this end, we will use the identity \eqref{eq42i4}.
Choosing $\Delta = \Delta_3$, $P^A = P_1^{A_1}$, and $\mathcal{F}_{\Delta_1 -1} (P_1) = \prod_{i<j} ( P_{ij})^{-\delta_{ij}+\frac{1}{2}} $, we obtain
\be \left[ \left( \delta_{12} - \frac{1}{2} \right) \frac{P_2^{A_1}}{P_{12}} +  \left( \delta_{13} - \frac{1}{2} \right) \frac{P_3^{A_1}}{P_{13}} \right]  \prod_{i<j} ( P_{ij})^{-\delta_{ij}+\frac{1}{2}}= 0 \ee
Therefore,
\be \mathcal{I} (2,3,1) = \frac{2\eta_{A_2A_3}P_{2A_1} }{ (\delta_{13} -\frac{1}{2})P_{12}P_{23}}   \ee
up to terms which vanish upon acting with $D^{M_1A_1}$, and similarly for $\mathcal{I} (3,1,2)$.

There are more terms in the amplitude involving $P_3^{A}$, due to the action of $D^{M_3A_3}$ on the off-shell leg, which also need to be eliminated.
We have
\be\label{eq37c} P_{3A_3}\mathcal{I} (1,2,3) + \dots + \dots =  \frac{1}{(\delta_{13} -\frac{1}{2})P_{12}} \left( - \eta^{A_1A_2} + \frac{2P_2^{A_1}P_3^{A_2} }{ P_{23}} \right) - ( 1\leftrightarrow 2) \ee
Using the identity \eqref{eq42i4} again, the second term on the right-hand side of \eqref{eq37c} is easily seen to be symmetric, and therefore vanishes. We arrive at an expression which is independent of $P_3$,
\be\label{eq37cx} P_{3A_3}\mathcal{I} (1,2,3) + \dots + \dots = \frac{\Delta_1 -\Delta_2}{(\delta_{23} -\frac{1}{2})(\delta_{13} -\frac{1}{2})} \frac{ \eta^{A_1A_2}}{P_{12}}  \ee
Notice that in the case of $\Delta_1 = \Delta_2$ (on-shell legs), this vanishes, so the action of $D^{M_3A_3}$ is simple in this case.

Differentiating with respect to $P_3$, we obtain an additional factor,
\be \frac{\partial}{\partial P_3^{M_3}} \prod_{i<j} ( P_{ij})^{-\delta_{ij}+\frac{1}{2}} = \left[ (2\delta_{13} -1) \frac{P_1^{M_3}}{P_{13}} + (2\delta_{23} -1) \frac{P_2^{M_3}}{P_{23}} \right] \prod_{i<j} ( P_{ij})^{-\delta_{ij}+\frac{1}{2}} \ee
It follows that
\be\label{eq48} D_{M_3A_3}\mathcal{A}^{A_1A_2A_3} = \frac{\pi^{d/2}}{2}\Gamma \left( \frac{\Delta_{1}+\Delta_{2}+\Delta_{3} -d+1}{2}  \right) \left[ (\mathcal{D}_3\mathcal{I}) (1,2) - (\mathcal{D}_3\mathcal{I}) (2,1) \right] \prod_{i<j} \Gamma\left( \delta_{ij} +\frac{1}{2} \right) ( P_{ij})^{-\delta_{ij}+\frac{1}{2}}\ee
where
\be (\mathcal{D}_3\mathcal{I}) (1,2) = \frac{(\Delta_3+ \Delta_1- \Delta_2 -1) \eta^{A_1A_2} P_{1M_3} - 2(\Delta_3-1) 
\delta^{A_1}_{M_3} P_1^{A_2}}{ \Delta_3 (\delta_{23} -\frac{1}{2})P_{12}P_{13}} \ee
In this form, the three-point vector amplitude can be used in higher-point amplitudes in much the same way as its scalar counterpart \eqref{eq1f}.

\section{Four-point Amplitudes}
\label{secIII}

In this section, we calculate scalar and vector four-point amplitudes by sewing together two three-point amplitudes calculated in section \ref{secII}. Using the results in appendix \ref{Xintegral}, the integrals over AdS space are performed with little effort. In the vector case, there is an additional type of diagram contributing due to the existence of a four-point vertex. A quartic interaction can also be added in the scalar case. The calculation proceeds as in the vector case.

\subsection{Scalar amplitudes}

The four-point scalar amplitude reads (Fig.\ \ref{4pt})
\be\label{eqA4s0} A^{(4,s)} (\Delta_1,P_1;\Delta_2,P_2;\Delta_3,P_3;\Delta_4,P_4) = \frac{g^2}{\prod_i 2\pi^{d/2} \Gamma (\Delta_i +1 -\frac{d}{2})} \int \frac{dc}{2\pi i} f_{\delta,0} (c) \mathcal{A}_{4} (\Delta_1, P_1; \Delta_2, P_2; \Delta_3, P_3; \Delta_4, P_4 | c) \ee
where
\be\label{eqA4s}  \mathcal{A}_{4} (\Delta_1, P_1; \Delta_2, P_2; \Delta_3, P_3; \Delta_4, P_4 | c) = \int_{\partial AdS} dQ \mathcal{A}_{3}(\Delta_1,P_1;\Delta_2,P_2;d/2+c,Q) \mathcal{A}_{3}(\Delta_3,P_3;\Delta_4,P_4; d/2-c, Q) \ee
and $\mathcal{A}_{3}$ is given by (\ref{eq1f}). 
\tikzset{
particle/.style={thick,draw=black},
particle2/.style={dotted,thick,draw=black
 }}
\begin{center}
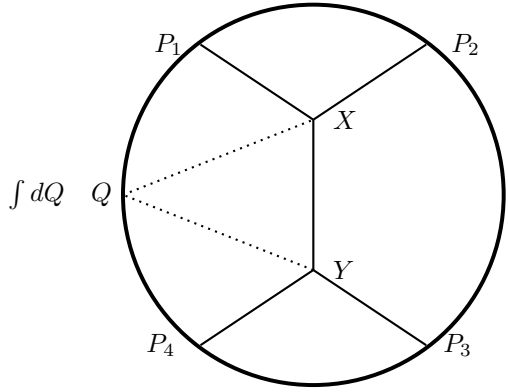

\begin{tikzpicture}[node distance=1cm and 1.5cm]
\coordinate[label={[xshift=-3pt]left:$P_1$}] (e1);
\coordinate[below right=of e1, label={[xshift=-6pt]right:$~~~X$}] (aux1);
\coordinate[above right=of aux1,label={[xshift=6pt]right:$P_2$}] (e2);
\coordinate[below=2cm of aux1, label={[xshift=-6pt]right:$~~~Y$}] (aux2);
\coordinate[below left=of aux2,label={[xshift=-6pt]left:$P_4$}] (e3);
\coordinate[below right=of aux2,label={[xshift=3pt]right:$P_3$}] (e4);

\draw[particle] (aux2) -- (aux1);
\draw[particle] (e1) -- (aux1);
\draw[particle] (aux1) -- (e2);
\draw[particle] (e3) -- (aux2);
\draw[particle] (aux2) -- (e4);
\node[draw,name path=circle,line width=1.5pt,circle,fit=(e1) (e4),inner sep=.5\pgflinewidth] {};
\path[name path=diameter] let \p1=(aux1), \p2=(aux2) 
  in (aux1|-0,0.5*\y2+0.5*\y1) -- ++(-3cm,0);
\path[name intersections={of=circle and diameter, by={aux3}}];
\draw[particle2] (aux2) -- (aux3);
\draw[particle2] (aux3) -- (aux1);
\node[label={[xshift=3pt]left: $\int dQ ~~~Q$ }] at (aux3) {};
\end{tikzpicture}
\captionof{figure}{The four-point scalar amplitude \eqref{eqA4s0}.}
\label{4pt}
\end{center}
To integrate over $Q$, we need to calculate
\be \int_{\partial AdS} dQ \prod_{i=1}^4 \Gamma (\lambda_i) (-2Q\cdot P_i)^{-\lambda_i}~, \ee
where
\be \lambda_1 = \frac{\Delta_1-\Delta_2 + d/2+c}{2} \ , \ \ \lambda_2 = \frac{\Delta_2 - \Delta_1 + d/2+c}{2} \ , \ \
\lambda_3 = \frac{\Delta_3-\Delta_4 +d/2-c}{2} \ , \ \ \lambda_4 = \frac{\Delta_4-\Delta_3 +d/2-c}{2} \ee
Notice that $\lambda_1 + \dots + \lambda_4 =d$.
Using the result \eqref{eqA10} in the Appendix, we obtain
\be \int_{\partial AdS} dQ \prod_{i=1}^4\Gamma (\lambda_i) (-2Q\cdot P_i)^{-\lambda_i} = \frac{\pi^{d/2}}{2} \int \prod_{i<j} d\tilde\delta_{ij} \Gamma (\tilde\delta_{ij}) P_{ij}^{-\tilde\delta_{ij}}\ee
where the integration variables are constrained by
\be\label{eq24} \sum_{j\ne i} \tilde\delta_{ij} = \lambda_i \ee
The integration variables are related to the Mandelstam invariants by
\be \delta_{12} = \frac{\Delta_1+\Delta_2 - d/2-c}{2} + \tilde\delta_{12} \ \ , \ \ \ \ \delta_{34} = \frac{\Delta_3+\Delta_4 -d/2+c}{2} + \tilde\delta_{34}~, \ee
and $\delta_{ij} = \tilde\delta_{ij}$, otherwise.
The constraints \eqref{eq24} in terms of the standard Mandelstam variables read
\be\label{eqMan} \sum_{j\ne i} \delta_{ij} = \Delta_i \ee
as expected (Eq.\ \eqref{eqMan}).

The four-point function \eqref{eqA4s} becomes
%
\be\label{eq25a} \mathcal{A}_{4} (\{\Delta_i,P_i\} | c) = \frac{\pi^{d/2}}{2} \int \prod_{i<j} d\delta_{ij} \Gamma (\delta_{ij})  \mathcal{M}_4 (\delta_{ij} | c) P_{ij}^{-\delta_{ij}} \ee
where
\be\label{eq31} \mathcal{M}_4 = \frac{\prod_{\sigma=\pm}\Gamma (\frac{\Delta_1+\Delta_2 -d/2+\sigma c}{2})\Gamma (\frac{\Delta_3+\Delta_4 -d/2+\sigma c}{2})\Gamma (\delta_{12} -  \frac{\Delta_1+\Delta_2 - d/2+\sigma c}{2}) }{\Gamma (\delta_{12})\Gamma (\delta_{34}) } \ee
Notice that the Mellin transform \eqref{eq31} is a function of a single Mandelstam invariant, $\delta_{12}$, because $\delta_{34}$ and $\delta_{12}$ are related through $\delta_{12} - \delta_{34} = \frac{\Delta_{1} + \Delta_2 - \Delta_3 - \Delta_4}{2}$.


\subsection{Vector amplitudes}

In the vector case, there are two types of diagrams, and we consider them separately.
First we discuss the four-point diagram due to the four-point gauge interaction (Fig.\ \ref{4ptv1}). The amplitude is
\be\label{eq20c}
A^{(4,v,(a))M_1 M_2 M_3 M_4,a_1a_2a_3a_4}=\int_{AdS} dX\, I_{A_1A_2A_3A_4}^{a_1a_2a_3a_4} \prod_{i=1}^4 E^{M_iA_i} (X,P_i)~,
\ee
where
\be I_{A_1A_2A_3A_4}^{a_1a_2a_3a_4} = \left( f^{a_1a_4b} f^{a_2a_3b} + f^{a_1a_3b} f^{a_2a_4b} \right) \eta_{A_1 A_2} \eta_{A_3 A_4} + \cdots + \cdots \ee
is independent of the points $P_i$ ($i=1,2,3,4$) and $X$.

\tikzset{
particle/.style={decorate, draw=black,
    decoration={coil,aspect=0.08,segment length=3pt,amplitude=3pt}}}
\begin{center}
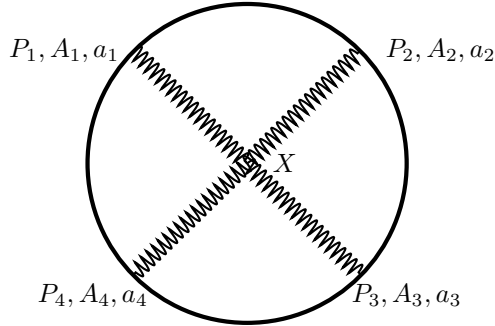

\begin{tikzpicture}[thick, node distance=1.5cm and 1.5cm]
\coordinate[label={[xshift=-2pt]left:$P_1, A_1, a_1$}] (e1);
\coordinate[below right=of e1, ,label={[xshift=6pt]right:$X$}] (aux1);
\coordinate[above right=of aux1,label={[xshift=6pt]right:$P_2,A_2,a_2$}] (e2);
\coordinate[below=1/cos(45)*2.1cm of e1,label={[xshift=6pt]below:$P_4, A_4, a_4~~~~~~~~~~~~~$}] (aux2);
\coordinate[below=1/cos(45)*2.1cm of e2,label={[xshift=6pt]below:$~~~~~~~P_3, A_3, a_3$}] (aux3);
\draw[particle] (e1) -- (aux1);
\draw[particle] (aux1) -- (e2);

\draw[particle] (aux1) --  (aux3);
\draw[particle] (aux1) --  (aux2);
\draw  [ultra thick] (aux1) circle [radius=sqrt(1.5cm^2+1.5cm^2)];
\end{tikzpicture}
\captionof{figure}{The four-point vector amplitude \eqref{eq20c}.}
\label{4ptv1}
\end{center}

The  integral over the bulk vector $X$ is of the form \eqref{eqA12}. Using the result \eqref{eqA16}, we obtain 
\be\label{eq44}
A^{(4,v,(a))M_1 M_2 M_3 M_4,a_1a_2a_3a_4} = \frac{\pi^{d/2}}{2\prod_i 2\pi^{d/2} \Gamma (1+\Delta_i -\frac{d}{2})} \prod_{i=1}^4 D^{M_i A_i} (\Delta_i, P_i) \mathcal{A}_{A_1A_2A_3A_4}^{a_1a_2a_3a_4} \ee
where
\be  \mathcal{A}_{A_1A_2A_3A_4}^{a_1a_2a_3a_4}=\int \mathcal{M}_{A_1A_2A_3A_4}^{a_1a_2a_3a_4} (\delta_{ij})\prod_{i < j} \Gamma (\delta_{ij}) P_{ij}^{-\delta_{ij}} d\delta_{ij}
\ \ , \ \ \ \  \mathcal{M}_{A_1A_2A_3A_4}^{a_1a_2a_3a_4} = \Gamma \left( \frac{\sum_i \Delta_i -d}{2} \right) I_{A_1A_2A_3A_4}^{a_1a_2a_3a_4}\ee
On shell ($\Delta_i = d-1$, $i=1,2,3,4$), this amplitude reads
\be  \mathcal{A}_{A_1A_2A_3A_4}^{a_1a_2a_3a_4}= \Gamma \left( \frac{3d-4}{2} \right) I_{A_1A_2A_3A_4}^{a_1a_2a_3a_4}\int\prod_{i < j} \Gamma (\delta_{ij}) P_{ij}^{-\delta_{ij}} d\delta_{ij}~. \ee
To use it in a higher-point diagram, we need to act with $D^{M_4A_4}$ and eliminate $P_4$ with a free index.
By using the identity \eqref{eq42i4} with $\Delta = \Delta_3$, $P^A = P_3^{A_3}$, and $\mathcal{F}_{\Delta_3-1} (P_3) = P_{34} \prod_{i<j} P_{ij}^{-\delta_{ij}}$, we obtain
\be\label{eq73} \left[ \delta_{13} \frac{P_{34}}{P_{13}} P_1^{A_3} +  \delta_{23} \frac{P_{34}}{P_{23}} P_2^{A_3} +  (\delta_{34} -1) P_4^{A_3} \right] \prod_{i<j}  P_{ij}^{-\delta_{ij}} = 0~,\ee
up to terms which vanish upon acting with $D_{M_3A_3}$.
We deduce
\be P_4^{A_4}  \eta_{A_3 A_4} = -\frac{1}{\delta_{34} -1}   \left[ \delta_{13} \frac{P_{1A_3} }{P_{13}} +  \delta_{23} \frac{P_{2A_3}}{P_{23}}  \right] P_{34} \ee
Differentiation with respect to $P_{4M_4}$ has the effect of multiplication by a factor given by
\be \frac{\partial}{\partial P_{4M_4}} P_{34} \prod_{i=1}^3 P_{i4}^{-\delta_{i4}} = 2  \left[  \delta_{14}\frac{P_{1}^{M_4}}{P_{14}} + \delta_{24}\frac{P_{2}^{M_4}}{P_{24}} + (\delta_{34} -1) \frac{P_{3}^{M_4}}{P_{34}} \right]P_{34} \prod_{i=1}^3 P_{i4}^{-\delta_{i4}}\ee
It follows that in the amplitude \eqref{eq44},
\be D^{M_4A_4} \eta_{A_3 A_4} =  \frac{\Delta_4 -1}{\Delta_4}\delta_{A_3}^{ M_4} -  \frac{2 }{\Delta_4} \left( \delta_{13} \frac{P_{1A_3} }{P_{13}} +  \delta_{23} \frac{P_{2A_3}}{P_{23}}  \right)  \left(  \frac{\delta_{14}}{\delta_{34} -1}  \frac{P_{34}}{P_{14}} P_{1}^{M_4}+ \frac{\delta_{24}}{\delta_{34} -1}  \frac{P_{34}}{P_{24}} P_{2}^{M_4}+ P_{3}^{M_4} \right) \ee
This expression allows us to use this amplitude in the calculation of a higher-point amplitude in which the leg corresponding to $P_4$ is internal.

\tikzset{
particle/.style={decorate, draw=black,
    decoration={coil,aspect=0.08,segment length=3pt,amplitude=3pt}}}
\tikzset{
particle2/.style={densely dotted  ,decorate, draw=black,
    decoration={coil,aspect=0.08,segment length=3pt,amplitude=3pt}}}

\begin{center}
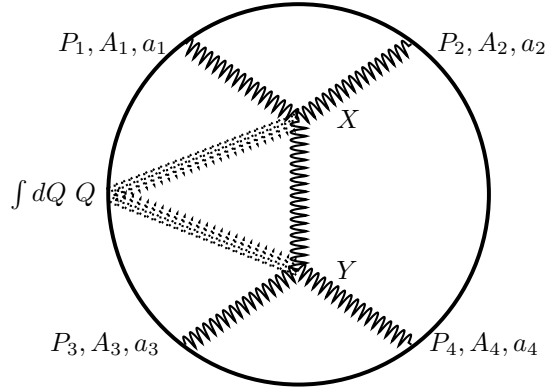

\begin{tikzpicture}[thick,node distance=1cm and 1.5cm] 
\coordinate[label={[xshift=-3pt]left:$P_1, {A_1}, a_1$}] (e1);
\coordinate[below right=of e1, label={[xshift=-6pt]right:$~~~~~X$}] (aux1);
\coordinate[above right=of aux1,label={[xshift=6pt]right:$P_2, {A_2}, a_2$}] (e2);
\coordinate[below=2cm of aux1, label={[xshift=-6pt]right:$~~~~~Y$}] (aux2);
\coordinate[below left=of aux2,label={[xshift=-6pt]left:$P_3, {A_3}, a_3$}] (e3);
\coordinate[below right=of aux2,label={[xshift=3pt]right:$P_4, {A_4}, a_4$}] (e4);

\draw[particle] (aux2) -- (aux1);
\draw[particle] (e1) -- (aux1);
\draw[particle] (aux1) -- (e2);
\draw[particle] (e3) -- (aux2);
\draw[particle] (aux2) -- (e4);
\node[draw,name path=circle,line width=1.5pt,circle,fit=(e1) (e4),inner sep=.5\pgflinewidth] {};
\path[name path=diameter] let \p1=(aux1), \p2=(aux2) 
  in (aux1|-0,0.5*\y2+0.5*\y1) -- ++(-3cm,0);
\path[name intersections={of=circle and diameter, by={aux3}}];
\draw[particle2] (aux2) -- (aux3);
\draw[particle2] (aux3) -- (aux1);
\node[label={[xshift=3pt]left: $\int dQ ~Q$ }] at (aux3) {};
\end{tikzpicture}
\captionof{figure}{The four-point vector amplitude \eqref{eq57o}.}
\label{3ptv}
\end{center}
Next, we calculate the four-point vector amplitude depicted in Fig.\ \ref{3ptv}. We have
\be\label{eq57o}
A^{(4,v, (b))M_1 M_2 M_3 M_4,a_1a_2a_3a_4}= g^2 f^{a_1a_2b} f^{a_3a_4b} \int \frac{dc}{2\pi i} f_{\delta,1} (c)  \prod_{i=1}^4 D^{M_iA_i} (\Delta_i,P_i) \mathcal{A}_{A_1 A_2 A_3A_4} (\{ \Delta_i, P_i \} |c ) \ee
where
\bes\label{eq57} \mathcal{A}_{A_1 A_2 A_3A_4} (\{ \Delta_i, P_i \} |c ) &=& \int_{\partial AdS} dQ 
 \eta_{NN'}D^{NC} (d/2+c, Q)\mathcal{A}_{A_1A_2C} (\Delta_1, P_1; \Delta_2, P_2; d/2+c, Q) \nonumber\\
&&\times D^{N' C'} (d/2-c,Q)\mathcal{A}_{A_3 A_4 C'} (\Delta_3,P_3;\Delta_4, P_4; d/2-c, Q)
\ees
Using \eqref{eq48}, we can express this in terms of  the scalar functions \eqref{eqA4s}. The integral over $Q$ corresponding to the product of \eqref{eq48} and its counterpart in the second amplitude (with $1\to 3$ and $2\to 4$) is
\be \int_{\partial AdS} dQ \prod_{i=1}^4 \Gamma (\lambda_i)(-2P_i\cdot Q)^{-\lambda_i} = \frac{\pi^{d/2}}{2} \int \prod_{i<j} d\tilde\delta_{ij} \Gamma (\tilde\delta_{ij}) P_{ij}^{-\tilde\delta_{ij}} \ee
where
\be \lambda_1 = \frac{\Delta_1-\Delta_2 + \frac{d}{2} +c+1}{2} \ , \ \ \lambda_2 = \frac{\Delta_2 - \Delta_1 +  \frac{d}{2}+c-1}{2} \ , \ \
\lambda_3 = \frac{\Delta_3 - \Delta_4 +  \frac{d}{2}-c+1}{2} \ , \ \ \lambda_4 = \frac{\Delta_4 - \Delta_3 +  \frac{d}{2}-c-1}{2} \ee
The Mandelstam invariants are
\be \delta_{12} = \tilde\delta_{12} + \frac{\Delta_1+\Delta_2 -  \frac{d}{2} -c+1}{2} \ , \ \ \delta_{34} = \tilde\delta_{34} + \frac{\Delta_3-\Delta_4 - \frac{d}{2} +c+1}{2}  \ , \ \ \delta_{13} = \tilde\delta_{13} -1~, \ee
and $\delta_{ij} = \tilde\delta_{ij}$, otherwise.
Thus, the four-point vector amplitude \eqref{eq57} can be put in the form \eqref{eq25a}, as in the scalar case,
\be\label{eq25av} \mathcal{A}_{A_1 A_2 A_3A_4} (\{\Delta_i,P_i\} | c) = \frac{\pi^{d/2}}{2} \int  \mathcal{M}_{A_1 A_2 A_3A_4} (\delta_{ij} | c) \prod_{i<j} \Gamma (\delta_{ij}) P_{ij}^{-\delta_{ij}} d\delta_{ij} \ee
where
\be \mathcal{M}_{A_1 A_2 A_3A_4} (\delta_{ij} | c)= \frac{(\frac{d}{2}-1)^2 -c^2}{\frac{d^2}{4}-c^2} \left[ \delta_{13}\mathcal{I}({1,2,3,4}) -\delta_{14}\mathcal{I}({1,2,4,3}) -\delta_{23}\mathcal{I}({2,1,3,4}) +\delta_{24}\mathcal{I}({2,1,4,3}) \right]
\mathcal{M}_{4} ~,\ee
$\mathcal{M}_4$ is as in the scalar case (Eq.\ \eqref{eq31}), but with the replacements $\Delta_1\to \Delta_1+1,\Delta_3\to\Delta_3+1$,
and
\bes \mathcal{I}({1,2,3,4})&=& -\frac{ (\frac{d}{2}+c- \Delta_1+ \Delta_2 -1) (\frac{d}{2}-c- \Delta_3+ \Delta_4 -1) }{2[(\frac{d}{2}-1)^2 -c^2]} \eta_{A_1A_2}\eta_{A_3A_4} \nextl
- 2\frac{\frac{d}{2}+c- \Delta_1+ \Delta_2 -1}{ \frac{d}{2}+c-1} \eta_{A_1A_2} \frac{P_{1A_3} P_{3A_4}}{P_{13}}
-2\frac{\frac{d}{2}-c- \Delta_3+ \Delta_4 -1}{\frac{d}{2}-c-1} \eta^{A_3A_4} 
\frac{P_{1A_2} P_{3A_1}}{P_{13}} \nextl
+4
\eta_{A_1A_3} \frac{P_{1A_2} P_{3A_4}}{P_{13}} 
\ees 
The above expressions simplify for on-shell amplitudes. Setting $\Delta_i = d-1$ ($i=1,2,3,4$), we obtain
\bes \mathcal{M}_{A_1 A_2 A_3A_4} (\delta_{ij} | c) &=& \frac{(\frac{d}{2}-1)^2 -c^2}{\frac{d^2}{4}-c^2} \left[  -  \eta_{A_1A_2}\eta_{A_3A_4}
- 2 \eta_{A_1A_2}\left( \frac{P_{1A_3} P_{3A_4}}{P_{13}} + \frac{P_{2A_4} P_{4A_3}}{P_{24}}\right) \right. \nextl
\left. -2 \eta_{A_3A_4} \left( 
\frac{P_{1A_2} P_{3A_1}}{P_{13}}+
\frac{P_{2A_1} P_{4A_2}}{P_{24}} \right)
+4
\eta_{A_1A_3} \frac{P_{1A_2} P_{3A_4}}{P_{13}} +4
\eta_{A_2A_4} \frac{P_{2A_1} P_{4A_3}}{P_{24}} \right] 
\delta_{13} \mathcal{M}_{4} \nextl - (3\longleftrightarrow 4) ~,\ees
where
\be\label{eq31von} \mathcal{M}_4 = \frac{\prod_{\sigma=\pm}\Gamma^2 (\frac{\frac{3d}{2}-1 +\sigma c}{2})\Gamma (\delta_{12} -  \frac{\frac{3d}{2} - 1+\sigma c}{2}) }{\Gamma^2 (\delta_{12}) } \ee
To use this function in higher-point amplitudes, it is convenient to eliminate all terms with $P_4^{A_i}$ ($i=1,2,3$).

By using the identity \eqref{eq42i4} with $\Delta = \Delta_1$, $P^A = P_1^{A_2}$, and $\mathcal{F}_{\Delta_1} (P_1) = P_{1}^{A_2} \prod_{i<j} P_{ij}^{-\delta_{ij}}$, we obtain
\be\label{eq73v} \left[  \frac{1}{2}\eta^{A_1A_2} + \sum_{k\ne 1}\delta_{1k} \frac{P_k^{A_1}P_1^{A_2}}{P_{1k}} \right] \prod_{i<j} P_{ij}^{-\delta_{ij}} = 0\ee
We deduce from \eqref{eq73} and \eqref{eq73v},
\bes \mathcal{I}({1, 2, 4,3}) &=&  -\frac{ (\frac{d}{2}+c- \Delta_1+ \Delta_2 -1) (\frac{d}{2}-c+ \Delta_3- \Delta_4 -1) }{2[(\frac{d}{2}-1)^2 -c^2]} \eta_{A_1A_2}\eta_{A_3A_4} \nextl
-\frac{2}{\delta_{34} -1}\left( 2
\eta_{A_1A_4} P_{1A_2} - \frac{\frac{d}{2}+c- \Delta_1+ \Delta_2 -1}{ \frac{d}{2}+c-1} \eta_{A_1A_2} P_{1A_4} \right) \left( \delta_{13} \frac{ P_{1A_3}}{P_{13}} +  \delta_{23} \frac{ P_{2A_3}}{P_{23}} \right)\frac{P_{34}}{P_{14}}\nextl
+\frac{2}{\delta_{14}}\frac{\frac{d}{2}-c- \Delta_3+ \Delta_4 -1}{\frac{d}{2}-c-1} \eta^{A_3A_4} 
\left( \frac{1}{2}\eta^{A_1A_2} + \delta_{12} \frac{P_{2A_1}P_{1A_2}}{P_{12}} + \delta_{13} \frac{P_{3A_1}P_{1A_2}}{P_{13}} \right)
\ees 
and similarly for $\mathcal{I}({2,1,4,3})$. $\mathcal{I}({1,2,3,4})$ and $\mathcal{I}({2,1,3,4})$ are unaltered.

Next we act with $D^{M_4A_4}$. We have
\bes\label{eq88} P_{4}^{A_4}\mathcal{I}({1 ,2, 3,4}) &=&  -\frac{ (\frac{d}{2}+c- \Delta_1+ \Delta_2 -1) \frac{d}{2}-c- \Delta_3+ \Delta_4 -1) }{2[(\frac{d}{2}-1)^2 -c^2]} \eta_{A_1A_2} P_{4A_3} \nextl
+\frac{\frac{d}{2}+c- \Delta_1+ \Delta_2 -1}{ \frac{d}{2}+c-1} \eta_{A_1A_2} P_{1A_3} \frac{P_{34}}{P_{13}}
-2\frac{\frac{d}{2}-c- \Delta_3+ \Delta_4 -1}{\frac{d}{2}-c-1} \frac{P_{4A_3} 
P_{1A_2} P_{3A_1}}{P_{13}} \nextl
+4
\eta_{A_1A_3} P_{1A_2} \frac{P_{34}}{P_{13}}\ees
and
\bes\label{eq89}
P_{4}^{A_4}\mathcal{I}({1 ,2, 4,3}) &=& -\frac{ (\frac{d}{2}+c- \Delta_1+ \Delta_2 -1) (\frac{d}{2}-c+ \Delta_3- \Delta_4 -1) }{2[(\frac{d}{2}-1)^2 -c^2]} \eta_{A_1A_2}P_{4A_3} \nextl
-\frac{1}{\delta_{34} -1}\left( 4
P_{4A_1} P_{1A_2} + \frac{\frac{d}{2}+c- \Delta_1+ \Delta_2 -1}{ \frac{d}{2}+c-1} \eta_{A_1A_2} P_{14} \right) \left( \delta_{13} \frac{ P_{1A_3}}{P_{13}} +  \delta_{23} \frac{ P_{2A_3}}{P_{23}} \right)\frac{P_{34}}{P_{14}}\nextl
+\frac{2}{\delta_{14}}\frac{\frac{d}{2}-c- \Delta_3+ \Delta_4 -1}{\frac{d}{2}-c-1} P_{4A_3} 
\left( \eta_{A_1A_2} + \delta_{12} \frac{P_{2A_1}P_{1A_2}}{P_{12}} + \delta_{13} \frac{P_{3A_1}P_{1A_2}}{P_{13}} \right)
\ees 
and similarly for $P_{4}^{A_4}\mathcal{I}({2,1,3,4})$ and $P_{4}^{A_4}\mathcal{I}({2,1,4,3})$.

$P_4$ with a free index is eliminated from \eqref{eq88} using \eqref{eq73} and
\be\label{eq73v2} \left[  \frac{1}{2}\eta^{A_1A_3} + (\delta_{13} +1)\frac{P_3^{A_1}P_1^{A_3}}{P_{13}} + \delta_{23} \frac{P_3^{A_1}P_2^{A_3}}{P_{23}} + (\delta_{34} -1) \frac{P_3^{A_1}P_4^{A_3}}{P_{34}} \right] \frac{P_{34}}{P_{13}}\prod_{i<j} P_{ij}^{-\delta_{ij}} = 0\ee
We obtain
\bes\label{eq88a} \frac{P_{13}}{P_{34}}P_{4}^{A_4}\mathcal{I}({1, 2, 3,4}) &=&  \frac{ (\frac{d}{2}+c- \Delta_1+ \Delta_2 -1) (\frac{d}{2}-c- \Delta_3+ \Delta_4 -1) }{2(\delta_{34} -1)[(\frac{d}{2}-1)^2 -c^2]} \eta_{A_1A_2}  \left( \delta_{13} P_{1A_3} +  \delta_{23} \frac{P_{13}}{P_{23}} P_{2A_3}  \right)\nextl
+\frac{\frac{d}{2}+c- \Delta_1+ \Delta_2 -1}{ \frac{d}{2}+c-1} \eta_{A_1A_2} P_{1A_3}
+4
\eta_{A_1A_3} P_{1A_2} \nextl
+\frac{2}{\delta_{34} -1}\frac{\frac{d}{2}-c- \Delta_3+ \Delta_4 -1}{\frac{d}{2}-c-1}
P_{1A_2}\left(  \frac{1}{2}\eta_{A_1A_3} + (\delta_{13} +1)\frac{P_{3A_1}P_{1A_3}}{P_{13}} + \delta_{23} \frac{P_{3A_1}P_{2A_3}}{P_{23}}  \right)
\ees
Notice that $P_4$ only enters through an overall factor of $P_{34}$.

Similarly, $P_4$ is eliminated from \eqref{eq89} using \eqref{eq73}, \eqref{eq73v}, \eqref{eq73v2}, and
\be\label{eq73v3} \left[  \frac{1}{2}\eta^{A_1A_2} P_1^{A_3}+\frac{1}{2}\eta^{A_1A_3} P_1^{A_2}+ \left( \delta_{12} \frac{P_2^{A_1}}{P_{12}} + (\delta_{13} +1) \frac{P_3^{A_1}}{P_{13}} + \delta_{14}  \frac{P_4^{A_1}}{P_{14}}\right) P_1^{A_2}P_1^{A_3} \right] \frac{1}{P_{13}}\prod_{i<j} P_{ij}^{-\delta_{ij}} = 0\ee
We obtain
\bes\label{eq89a}
\frac{P_{13}}{P_{34}}P_{4}^{A_4}\mathcal{I}({1, 2, 4,3}) &=& \frac{ (\frac{d}{2}+c- \Delta_1+ \Delta_2 -1) (\frac{d}{2}-c+ \Delta_3- \Delta_4 -1) }{2(\delta_{34} -1)[(\frac{d}{2}-1)^2 -c^2]} \eta_{A_1A_2} \left(\delta_{13}  P_{1A_3} +  \delta_{23} \frac{P_{13}}{P_{23}} P_{2A_3}  \right) \nextl
+\frac{4\delta_{13}}{\delta_{14}(\delta_{34} -1)}
   \left[  \frac{1}{2}\eta_{A_1A_2} P_{1A_3}+\frac{1}{2}\eta_{A_1A_3} P_{1A_2}+ \left( \delta_{12} \frac{P_{2A_1}}{P_{12}} + (\delta_{13} +1) \frac{P_{3A_1}}{P_{13}} \right) P_{1A_2}P_{1A_3} \right] \nextl
+\frac{4  \delta_{23}}{\delta_{14}(\delta_{34} -1)}
 \frac{ P_{2A_3}}{P_{23}}P_{13}\left( \frac{1}{2}\eta_{A_1A_2} + \delta_{12} \frac{P_{2A_1}P_{1A_2}}{P_{12}} + \delta_{13} \frac{P_{3A_1}P_{1A_2}}{P_{13}} \right)\nextl
-\frac{1}{\delta_{34} -1} \frac{\frac{d}{2}+c- \Delta_1+ \Delta_2 -1}{ \frac{d}{2}+c-1} \eta_{A_1A_2} \left( \delta_{13} P_{1A_3} +  \delta_{23} \frac{ P_{13}}{P_{23}} P_{2A_3} \right) \nextl
-\frac{2}{\delta_{14}(\delta_{34} -1)}\frac{\frac{d}{2}-c- \Delta_3+ \Delta_4 -1}{\frac{d}{2}-c-1}
\left( \eta_{A_1A_2} + \delta_{12} \frac{P_{2A_1}P_{1A_2}}{P_{12}}  \right) \left(\delta_{13}  P_{1A_3} +  \delta_{23} \frac{P_{13}}{P_{23}} P_{2A_3}  \right)\nextl
-\frac{2\delta_{13}}{\delta_{14}(\delta_{34} -1)}\frac{\frac{d}{2}-c- \Delta_3+ \Delta_4 -1}{\frac{d}{2}-c-1}
  P_{1A_2} \left(  \frac{1}{2}\eta_{A_1A_3} + (\delta_{13} +1)\frac{P_{3A_1}P_{1A_3}}{P_{13}} + \delta_{23} \frac{P_{3A_1}P_{2A_3}}{P_{23}}  \right)
\ees 
Again,  $P_4$ only enters through an overall factor of $P_{34}$. It follows that the part of the amplitude involving $P_4$ is $P_{34} \prod_{i=1}^3 P_{i4}^{-\delta_{i4}}$. Therefore, differentiation with respect to $P_{4M_4}$ has the effect of multiplication by an overall factor,
\be \frac{\partial}{\partial P_{4M_4}} P_{34} \prod_{i=1}^3 P_{i4}^{-\delta_{i4}} = 2  \left[  \delta_{14}\frac{P_{1}^{M_4}}{P_{14}} + \delta_{24}\frac{P_{2}^{M_4}}{P_{24}} + (\delta_{34} -1) \frac{P_{3}^{M_4}}{P_{34}} \right]P_{34} \prod_{i=1}^3 P_{i4}^{-\delta_{i4}}\ee
The amplitude with $D^{M_4A_4}$ acted upon is given by
\be\label{eq25av2} D^{M_4A_4}\mathcal{A}_{A_1 A_2 A_3A_4} (\{\Delta_i,P_i\} | c) = \frac{\pi^{d/2}}{2} \int (\mathcal{D}_4 \mathcal{M})^{M_4}_{A_1A_2A_3}  (\delta_{ij} | c)  \prod_{i<j}  P_{ij}^{-\delta_{ij}}  \Gamma (\delta_{ij}) d\delta_{ij} \ee
where
\be (\mathcal{D}_4 \mathcal{M})^{M_4}_{A_1A_2A_3} =\left[ \frac{\Delta_4 -1}{\Delta_4}\eta^{M_4A_4} + \frac{2}{\Delta_4}  \left(   \delta_{14}P_{1}^{M_4}\frac{P_{34}}{P_{14}} + \delta_{24}P_{2}^{M_4}\frac{P_{34}}{P_{24}} + (\delta_{34} -1) P_{3}^{M_4}\right) \frac{P_{4A_4}}{P_{34}} \right] \mathcal{M}_{A_1 A_2 A_3A_4} (\delta_{ij} | c)   \ee
In the above expression, the only dependence on $P_4$ is through the ratios $\frac{P_{34}}{P_{14}}$ and $\frac{P_{34}}{P_{24}}$.
This expression will be used in the calculation of five- and higher-point vector amplitudes.

\section{Five-point Amplitudes}
\label{secIV}

In this section, we calculate scalar and vector five-point amplitudes by sewing together three- and four-point amplitudes. The integrals over AdS space that are involved are similar to the ones encountered in the case of four-point amplitudes in section \ref{secIII} and can be performed using the results of appendix \ref{Xintegral} without additional effort. The Mellin amplitudes are found as integrals over the Mandelstam invariants of the constituent four-point amplitudes. These integrals can all be performed, resulting in expressions involving generalized Hypergeometric functions. Thus, our approach provides an alternative to integration over Schwinger parameters \cite{Paulos:2011ie}.

\subsection{Scalar amplitudes}

The five-point scalar amplitude (Fig.\ \ref{5pt}) reads
\be\label{eq77} A^{(5,s)} (\Delta_1,P_1;\dots;\Delta_5,P_5) = \frac{g^3}{\prod_i 2\pi^{d/2} \Gamma (\Delta_i + 1 - d/2)} \int \frac{dcdc'}{(2\pi i)^2} f_{\delta_1,0} (c) f_{\delta_2,0} (c') \mathcal{A}_{5}(\{\Delta_i,P_1\} |c,c') \ee
where
\be \mathcal{A}_{5}(\{\Delta_i,P_i \} |c,c') = \int_{\partial AdS} dQ
\mathcal{A}_{4}(\Delta_1,P_1;\Delta_2,P_2;\Delta_3, P_3; d/2+c',Q|c) \mathcal{A}_{3} (\Delta_4,P_4; \Delta_5, P_5;d/2-c',Q) \ee

\begin{center}
\tikzset{
particle/.style={thick,draw=black},
particle2/.style={dotted,thick,draw=black
 }}
\begin{tikzpicture}[thick, node distance=1cm and 1.5cm]
\coordinate[label={[xshift=-3pt]left:$P_1~$}] (e1);

\coordinate[below right=of e1,label={[xshift=3pt]right:$~X$}] (aux1);

\coordinate[above right=of aux1,label={[xshift=6pt]right:$~~P_2$}] (e2);

\coordinate[below=1cm of aux1] (aux);

\coordinate[below left=of aux2,label={[xshift=-6pt]left:$P_5~$}] (e3);

\coordinate[below right=of aux2,label={[xshift=3pt]right:$~~P_4$}] (e4);

\coordinate[below=1cm of aux1,label={[xshift=3pt]above left:$Z~~~$}] (aux3);

\coordinate[below=2cm of aux1,label={[xshift=3pt]right:$~Y$}] (aux2);

\draw[particle] (e1) -- (aux1);
\draw[particle] (aux1) -- (e2);
\draw[particle] (e3) -- (aux2);
\draw[particle] (aux2) -- (e4);
\draw[particle] (aux1) -- node {} (aux2);
\node[draw,name path=circle,line width=3pt,circle,fit=(e1) (e4),inner sep=.5\pgflinewidth] {};
\path[name path=diameter] let \p1=(aux1), \p2=(aux2) 
 in (aux1|-0,0.5*\y2+0.5*\y1) -- ++(3cm,0);
\path[name intersections={of=circle and diameter, by={aux3}}];

\path[name path=diameter] let \p1=(aux1), \p2=(aux2) 
 in (aux1|-0,0.5*\y2+0.5*\y1) -- ++(-3cm,0);
\path[name intersections={of=circle and diameter, by={aux4}}];

\draw[particle2] (aux) -- (aux4);
\draw[particle2] (aux2) -- (aux4);

\draw[particle] (aux) -- (aux3);
\node[label={[xshift=2pt]right:$P_3$}] at (aux3) {};
\node[label={[xshift=2pt]left:$\int dQ~~ Q$}] at (aux4) {};

\end{tikzpicture}

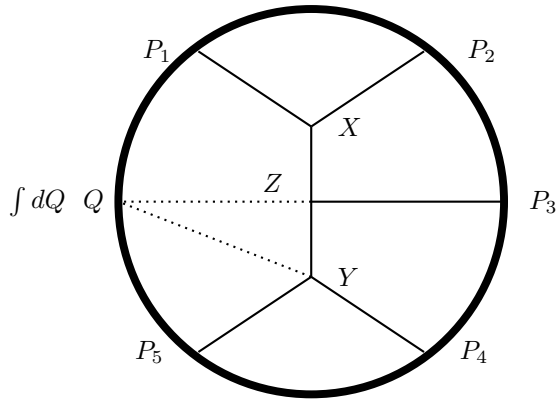
\captionof{figure}{The five-point scalar amplitude \eqref{eq77}.}
\label{5pt}
\end{center}

The integral over $Q$ involves
\be \int_{\partial AdS} dQ \prod_{i=1}^5\Gamma (\lambda_i) (-2Q\cdot P_i)^{-\lambda_i} \ee
where
\be \lambda_1 = \delta_{14}' \ , \ \ \lambda_2 = \delta_{24}' \ , \ \ \lambda_3 = \delta_{34}' \ , \ \ \lambda_4 = \frac{\Delta_4-\Delta_5 +\frac{d}{2}-c'}{2} \ , \ \ \lambda_5 = \frac{\Delta_5-\Delta_4 + \frac{d}{2}-c'}{2} \ee
and $\delta_{ij}'$ are the Mandelstam invariants for the four-point function constrained by
\bes \delta_{12}' + \delta_{13}' + \delta_{14}' &=& \Delta_1 \nonumber\\
\delta_{12}' + \delta_{23}' + \delta_{24}' &=& \Delta_2 \nonumber\\
\delta_{13}' + \delta_{23}' + \delta_{34}' &=& \Delta_3 \nonumber\\
\delta_{14}' + \delta_{24}' + \delta_{34}' &=& h+c'
\ees
Working as before, we obtain
\be \int_{\partial AdS} dQ \prod_{i=1}^5\Gamma (\lambda_i) (-2Q\cdot P_i)^{-\lambda_i} =  \frac{\pi^{d/2}}{2} \int \prod_{i<j} d\tilde\delta_{ij} \Gamma (\tilde\delta_{ij}) P_{ij}^{-\tilde\delta_{ij}} 
\ee
where the integration variables are constrained by
\be\label{eq245} \sum_{j\ne i} \tilde\delta_{ij} = \lambda_i \ee
The part of the amplitude involving the vectors $P_i$ is
\be P_{12}^{-\delta_{12}'} P_{13}^{-\delta_{13}'} P_{23}^{-\delta_{23}'} P_{45}^{- \frac{\Delta_4 + \Delta_5 - d/2 + c'}{2}}\prod_{i<j} P_{ij}^{-\tilde\delta_{ij}} = \prod_{i<j} P_{ij}^{-\delta_{ij}} \ee
where $\delta_{ij}$ are the Mandelstam invariants defined by
\be \delta_{12} = \tilde\delta_{12} + \delta_{12}' \ , \ \  \delta_{23} = \tilde\delta_{23} + \delta_{23}' \ , \ \ \delta_{13} = \tilde\delta_{13} + \delta_{13}' \ , \ \
\delta_{45} = \tilde\delta_{45} + \frac{\Delta_4 + \Delta_5 - d/2 + c'}{2} \ , \ee
and $\delta_{ij} = \tilde\delta_{ij}$, otherwise. It is easily seen that they obey the standard constraints \eqref{eqMan}.

The five-point function simplifies to
\be\label{eq25a5} \mathcal{A}_{5} (\{\Delta_i,P_i\} | c,c') = \frac{\pi^{d/2}}{2} \int \prod_{i<j} d\delta_{ij} \Gamma (\delta_{ij})  \mathcal{M}_5 (\delta_{ij} | c,c') P_{ij}^{-\delta_{ij}} \ee
where
\bes\label{eq43} \mathcal{M}_5 (\delta_{ij} | c,c') &=&\int \prod_{i<j} d\delta_{ij}' \frac{\Gamma (\delta_{12} - \delta_{12}')\Gamma (\delta_{12}')\Gamma (\delta_{23} - \delta_{23}')\Gamma (\delta_{23}') \Gamma(\delta_{13} - \delta_{13}')\Gamma (\delta_{13}')}{\Gamma(\delta_{12})\Gamma(\delta_{23})\Gamma(\delta_{13}) } \nextl
\times \frac{\Gamma (\delta_{45} - \frac{\Delta_4+\Delta_5 -d/2+c'}{2})\Gamma (\frac{\Delta_4+\Delta_5 -d/2+c'}{2})}{\Gamma (\delta_{45})} \mathcal{M}_4 (\delta_{12}' |c) \mathcal{M}_3 \ees
and
\be\mathcal{M}_3 = \Gamma \left( \frac{\Delta_4+\Delta_5 -\frac{d}{2} -c'}{2} \right) \ , \ \ \mathcal{M}_4 (\delta_{12}' |c) = \frac{\prod_{\sigma =\pm}\Gamma (\frac{\Delta_1+\Delta_2 -d/2+\sigma c}{2})\Gamma (\frac{\Delta_3+c' +\sigma c}{2})\Gamma (\delta_{12}' -  \frac{\Delta_1+\Delta_2 - d/2+\sigma c}{2}) }{\Gamma (\delta_{12}')\Gamma (\delta_{34}') } \ee
with $\delta_{34}' = \delta_{12}' - \frac{\Delta_1+\Delta_2-\Delta_3-d/2-c'}{2}$, $\delta_{23}' = \frac{\Delta_1 +\Delta_2 +\Delta_3 -d/2-c'}{2} - \delta_{12}' - \delta_{13}'$.

The two integrals in \eqref{eq43} are performed as follows. From Barnes first lemma, we have
\be \int \frac{d\delta_{13}'}{2\pi i} \frac{\Gamma(\delta_{23}') \Gamma (\delta_{23} - \delta_{23}')\Gamma (\delta_{13}') \Gamma (\delta_{13} - \delta_{13}')}{\Gamma (\delta_{13})\Gamma (\delta_{23})} =
\frac{\Gamma (\delta_{13} + \delta_{23} - \frac{\Delta_1+\Delta_2+\Delta_3-d/2-c'}{2} + \delta_{12}') \Gamma (\frac{\Delta_1+\Delta_2+\Delta_3 -d/2-c'}{2} - \delta_{12}')}{\Gamma (\delta_{13} + \delta_{23})}\ee
Next, we need
\bes \mathcal{F} &=& \int \frac{d\delta_{12}'}{2\pi i} \frac{\Gamma (\delta_{12}' + \delta_{13} + \delta_{23} - \frac{\Delta_1 + \Delta_2 + \Delta_3 -d/2-c'}{2} )\Gamma (\frac{\Delta_1+\Delta_2+\Delta_3 -d/2-c'}{2} - \delta_{12}') \Gamma (\delta_{12} - \delta_{12}') }{\Gamma (\delta_{13} + \delta_{23} )\Gamma (\delta_{12}' - \frac{\Delta_1+\Delta_2-\Delta_3-d/2-c'}{2}) }\nextl
\times  \prod_{\sigma = \pm}\Gamma \left( \delta_{12}' -  \frac{\Delta_1+\Delta_2 - \frac{d}{2}+\sigma c}{2} \right) \ees
This is calculated with the aid of the identity
\be \int \frac{ds}{2\pi i} \frac{\Gamma (a+s) \Gamma (b+s) \Gamma (f-c+s) \Gamma (e-a-b-s)\Gamma (-s)}{\Gamma (f+s)} =
\frac{\Gamma (a) \Gamma(b) \Gamma (e-a) \Gamma (e-b) \Gamma (f-c)}{\Gamma (e) \Gamma (f)} {}_3F_2 \left[ \begin{array}{c}a,b,c \\ e,f \end{array} \right] \ee
where ${}_3F_2 \left[ \begin{array}{c}a,b,c \\ e,f \end{array} \right] = {}_3F_2 ( a,b,c;  e,f ; 1)$.
We obtain
\bes \mathcal{F} &=& \frac{\Gamma (\delta_{45} - \frac{\Delta_4 + \Delta_5 -d/2-c'}{2} )  \Gamma (   \frac{ \Delta_3 +c-c'}{2} )\prod_{\sigma = \pm}\Gamma(\delta_{12} -  \frac{\Delta_1+\Delta_2 - \frac{d}{2}+\sigma c}{2})  }{\Gamma (\delta_{45} - \frac{\Delta_4 + \Delta_5-\Delta_3 -d/2-c}{2}) \Gamma (\delta_{12} - \frac{\Delta_1+\Delta_2-\Delta_3-d/2-c'}{2})} \nextl
\times {}_3F_2 \left[ \begin{array}{c} \delta_{45} - \frac{\Delta_4 + \Delta_5 -d/2-c'}{2}  ,\delta_{12} -  \frac{\Delta_1+\Delta_2 - \frac{d}{2}-c}{2},\frac{\Delta_3+c' + c}{2} \\
\delta_{45} - \frac{\Delta_4 + \Delta_5 -\Delta_3 -d/2-c}{2},\delta_{12} - \frac{\Delta_1+\Delta_2-\Delta_3-d/2-c'}{2} \end{array} \right]~. \ees
where we used $\delta_{12} + \delta_{13} + \delta_{23} - \frac{\Delta_1 + \Delta_2 }{2} = \delta_{45} - \frac{\Delta_4 + \Delta_5}{2}$.
Therefore,
\bes\label{eq43a} \mathcal{M}_5 (\delta_{ij} | c,c') &=& \frac{  \Gamma (   \frac{ \Delta_3 +c-c'}{2} )  \prod_{\sigma = \pm } \Gamma (\delta_{12} - \frac{\Delta_1+\Delta_2-d/2+\sigma c}{2})\Gamma(\delta_{45} -  \frac{\Delta_4+\Delta_5 - \frac{d}{2}+\sigma c'}{2})}{\Gamma (\delta_{12}) \Gamma (\delta_{12} - \frac{\Delta_1+\Delta_2-\Delta_3-d/2-c'}{2}) \Gamma (\delta_{45})\Gamma (\delta_{45} - \frac{\Delta_4 + \Delta_5 -\Delta_3 -d/2-c}{2})}\nextl
\times \prod_{\sigma = \pm}  \Gamma \left(   \frac{ \Delta_3 +\sigma c+c'}{2} \right) \Gamma\left( \frac{\Delta_1+\Delta_2 - \frac{d}{2}+\sigma c}{2} \right) \Gamma\left( \frac{\Delta_4+\Delta_5 - \frac{d}{2}+\sigma c'}{2} \right) \nextl
\times {}_3F_2 \left[ \begin{array}{c} \delta_{45} - \frac{\Delta_4 + \Delta_5 -d/2-c'}{2}  ,\delta_{12} -  \frac{\Delta_1+\Delta_2 - \frac{d}{2}-c}{2},\frac{\Delta_3+c' + c}{2} \\
\delta_{45} - \frac{\Delta_4 + \Delta_5 -\Delta_3 -d/2-c}{2},\delta_{12} - \frac{\Delta_1+\Delta_2-\Delta_3-d/2-c'}{2} \end{array} \right]~. \ees

\subsection{Vector amplitudes}

In order to avoid an unnecessarily long calculation, we restrict attention to the case of on-shell amplitudes by setting
$\Delta_i =d-1$ ($i=1,\dots, 5$).
There are two different diagrams which we need to consider separately.

The diagram depicted in figure \ref{5ptva} has amplitude
\be
\label{fivepointc}
A^{(5,v, (a))M_1 \dots M_5, a_1\dots a_5}= g^3\int \frac{dc}{2\pi i} f_{\delta,1} (c)  \prod_{i=1}^5 D^{M_iA_i} (\Delta_i,P_i) \mathcal{A}_{A_1 A_2 A_3A_4A_5}^{a_1a_2a_3a_4a_5} (\{ \Delta_i, P_i \} |c ) \ee
where
\bes\label{eq118a} \mathcal{A}_{A_1 A_2 A_3A_4A_5}^{a_1a_2a_3a_4a_5} (\{ \Delta_i, P_i \} |c ) &=& \int_{\partial AdS} dQ 
 \eta_{NN'}D^{NC} (h+c, Q) \mathcal{A}_{A_1A_2A_3C}^{a_1a_2a_3b} (\Delta_1, P_1; \Delta_2, P_2;\Delta_3 , P_3;  d/2+c, Q) \nonumber\\
&&\times D^{N' C'} (d/2-c,Q)f^{a_4a_5b} \mathcal{A}_{A_4 A_5 C'} (\Delta_4,P_4;\Delta_5, P_5; d/2-c, Q)
\ees
\begin{center}
\tikzset{
particle/.style={decorate, draw=black,
    decoration={coil,aspect=0.08,segment length=3pt,amplitude=3pt}}}\begin{tikzpicture}[thick, node distance=1cm and 1.5cm]
\coordinate[label={[xshift=-3pt]left:$P_1,{A_1}, a_1~~$}] (e1);

\coordinate[below right=of e1,label={[xshift=3pt]right:$~X$}] (aux1);

\coordinate[above right=of aux1,label={[xshift=6pt]right:$~~P_3,{A_3}, a_3$}] (e2);

\coordinate[below=1cm of aux1] (aux);
\coordinate[above=1.5cm of aux1] (auxabove);

\coordinate[below left=of aux2,label={[xshift=-6pt]left:$P_5, {A_5}, a_5~~$}] (e3);

\coordinate[below right=of aux2,label={[xshift=3pt]right:$~~P_4,{A_4}, a_4$}] (e4);


\coordinate[below=2cm of aux1,label={[xshift=3pt]right:$~Y$}] (aux2);

\draw[particle] (auxabove) -- (aux1);
\draw[particle] (e1) -- (aux1);
\draw[particle] (aux1) -- (e2);
\draw[particle] (e3) -- (aux2);
\draw[particle] (aux2) -- (e4);
\draw[particle] (aux1) -- node {} (aux2);
\node[draw,name path=circle,line width=3pt,circle,fit=(e1) (e4),inner sep=.5\pgflinewidth] {};
\path[name path=diameter] let \p1=(aux1), \p2=(aux2) 
 in (aux1|-0,0.5*\y2+0.5*\y1) -- ++(-3cm,0);
\path[name intersections={of=circle and diameter, by={aux3}}];

\path[name path=diameter] let \p1=(aux1), \p2=(aux2) 
 in (aux1|-0,0.5*\y2+0.5*\y1) -- ++(-3cm,0);
\path[name intersections={of=circle and diameter, by={aux4}}];

\draw[particle2] (aux1) -- (aux4);
\draw[particle2] (aux2) -- (aux4);

\node[label={[xshift=2pt]above:$P_2,{A_2},a_1$}] at (auxabove) {};
\node[label={[xshift=2pt]left:$\int dQ ~~Q$}] at (aux4) {};

\end{tikzpicture}

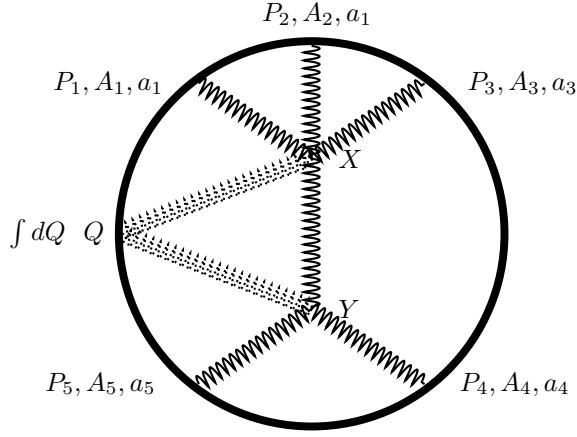
\captionof{figure}{The five-point vector amplitude \eqref{fivepointc}.}
\label{5ptva}
\end{center}
The three-point amplitude in \eqref{eq118a} simplifies to
\bes\label{eq48xa} D^{N' C'} (d/2-c,Q)\mathcal{A}_{A_4 A_5 C'} &=&\frac{2}{\frac{d}{2}-c} \Gamma \left( \frac{\frac{3d}{2}-c -1}{2}  \right)\Gamma \left( \frac{\frac{3d}{2}+c -1}{2}  \right) \Gamma^2 \left( \frac{\frac{d}{2}-c +1}{2}  \right)\nextl
\times \left[ \eta_{A_4A_5} P_{4}^{N'} - 2 
\delta_{A_4}^{N'} P_{4A_5} \right]  P_{45}^{-\frac{\frac{3d}{2}+c+1}{2}} ( -2P_4\cdot Q)^{-\frac{\frac{d}{2}-c+1}{2}} ( -2P_{5}\cdot Q)^{-\frac{\frac{d}{2}-c-1}{2}}\nextl
- (4\longleftrightarrow 5)\ees
The four-point amplitude in \eqref{eq118a} simplifies to
\be D^{NC} (d/2+c, Q) \mathcal{A}_{A_1A_2A_3C}^{a_1a_2a_3b} = \left( f^{a_1bb'} f^{a_2a_3b'} + f^{a_1a_3b'} f^{a_2bb'} \right) \int (\mathcal{D}_4 \mathcal{M})^{N}_{A_1A_2A_3}  (\delta_{ij}' | c)  \prod_{i<j}  P_{ij}^{-\delta_{ij}'}  \Gamma (\delta_{ij}') d\delta_{ij}' + \cdots + \cdots \ee
where
\bes (\mathcal{D}_4 \mathcal{M})^{N}_{A_1A_2A_3}  &=& \Gamma \left( \frac{\frac{5d}{2} -3 +c}{2} \right) \eta_{A_1A_2}
\left[ \frac{\frac{d}{2} +c -1}{\frac{d}{2} +c}\delta_{A_3}^{ N} \right. \nextl
\left. -  \frac{2 }{\frac{d}{2} +c} \left( \delta_{13}' \frac{P_{1A_3} }{P_{13}} +  \delta_{23}' \frac{P_{2A_3}}{P_{23}}  \right)  \left(  \frac{\delta_{14}'}{\delta_{34}' -1}  \frac{P_{3}\cdot Q}{P_{1}\cdot Q} P_{1}^{N}+ \frac{\delta_{24}'}{\delta_{34}' -1}  \frac{P_{3}\cdot Q}{P_{2}\cdot Q} P_{2}^{N}+ P_{3}^{N} \right) \right] \ees
The integral over $Q$ involves
\be\label{eq101} \int_{\partial AdS} dQ \prod_{i=1}^5\Gamma (\lambda_i) (-2Q\cdot P_i)^{-\lambda_i} = \frac{\pi^{d/2}}{2} \int \prod_{i<j} \frac{d\tilde\delta_{ij}}{2\pi i} \Gamma (\tilde\delta_{ij}) P_{ij}^{-\tilde\delta_{ij}}\ee
where
\be\label{eq102} \lambda_1 = \delta_{14}' +n_1\ , \ \ \lambda_2 = \delta_{24}'+n_2 \ , \ \ \lambda_3 = \delta_{34}' +n_3 \ , \ \ \lambda_4 = \frac{\frac{d}{2}-c+1}{2} \ , \ \ \lambda_5 = \frac{\frac{d}{2}-c-1}{2} \ee
and $\delta_{ij}'$ are the Mandelstam invariants for the four-point function, as in the scalar case. The various terms have $n_i \in \{-1,0,+1\}$, with $\sum_i n_i =0$ ($i=1,2,3$).
The integration variables are constrained by
\be\label{eq245n} \sum_{j\ne i} \tilde\delta_{ij} = \lambda_i \ee
In terms of the Mandelstam invariants,
\be\label{eq104} \tilde\delta_{12} = \delta_{12} - \delta_{12}' \ , \ \ \tilde\delta_{23} = \delta_{23} - \delta_{23}' \ , \ \ \tilde\delta_{13} = \delta_{13} - \delta_{13}' \ , \ \ \tilde\delta_{45} = \delta_{45} - \frac{\frac{3d}{2}+c-3}{2} \ , \ \ \tilde\delta_{i4} = \delta_{i4} + n_i \ \ \ (i=1,2,3)~, \ee
and $\tilde\delta_{ij} = \delta_{ij}$, otherwise.

We arrive at
\bes \mathcal{A}_{A_1 A_2 A_3A_4A_5} &=& \frac{\pi^{d/2}}{2} \int\prod_{i<j} \frac{d\tilde\delta_{ij}}{2\pi i} \Gamma (\tilde\delta_{ij}) P_{ij}^{-\tilde\delta_{ij}}\nextl
\times  \left[ \left( f^{a_1b'b} f^{a_2a_3b} + f^{a_1a_3b} f^{a_2b'b} \right) f^{b'a_4a_5} \mathcal{M}_{A_1\dots A_5} + \text{permutations of (123)} - (4\longleftrightarrow 5)  \right] \ees
where
\bes \mathcal{M}_{A_1\dots A_5} &=& \frac{\frac{d}{2}-c-1}{\frac{d}{2}-c} \Gamma \left( \frac{\frac{3d}{2}-c -1}{2}  \right)\Gamma \left( \frac{\frac{3d}{2}+c -1}{2}  \right) \Gamma \left( \frac{\frac{5d}{2} -3 +c}{2} \right) \eta_{A_1A_2}\left[ \eta_{A_4A_5} P_{4N} - 2 
\eta_{A_4N} P_{4A_5} \right] \nextl
\times \int\prod_{i<j}  d\delta_{ij}' \frac{\Gamma (\delta_{12} - \delta_{12}')\Gamma(\delta_{12}')\Gamma (\delta_{23} - \delta_{23}')\Gamma(\delta_{23}')\Gamma (\delta_{13} - \delta_{13}')\Gamma(\delta_{13}')\Gamma (\delta_{45} - \frac{\frac{3d}{2}+c-3}{2})}{\Gamma (\delta_{12})\Gamma (\delta_{23})\Gamma (\delta_{13})\Gamma (\delta_{45})}\nextl
\times \left[ \frac{\frac{d}{2} +c -1}{\frac{d}{2} +c}\delta_{A_3}^{ N} -  \frac{2 }{\frac{d}{2} +c} \left( \delta_{13}' \frac{P_{1A_3} }{P_{13}} +  \delta_{23}' \frac{P_{2A_3}}{P_{23}}  \right)  \left(  \frac{\delta_{14}}{\delta_{34} -1}   P_{1}^{N}+ \frac{\delta_{24}}{\delta_{34} -1}   P_{2}^{N}+ P_{3}^{N} \right) \right] \ees
The integrals over the four-point Mandelstam invariants $\delta_{ij}'$ can be performed as in the scalar case.

Next, we turn to the diagram depicted in figure \ref{5ptvb}. It is given by (suppressing standard group theory indices)
\be
\label{fivepoint}
A^{(5,v,(b))M_1 M_2 M_3 M_4M_5}= g^3 \int \frac{dcdc'}{(2\pi i)^2} f_{\delta,1} (c)  f_{\delta,1} (c')  \prod_{i=1}^5 D^{M_iA_i} (\Delta_i,P_i) \mathcal{A}_{A_1 A_2 A_3A_4A_5} (\{ \Delta_i, P_i \} |c,c' ) \ee
where
\bes\label{eq118} \mathcal{A}_{A_1 A_2 A_3A_4A_5} (\{ \Delta_i, P_i \} |c,c' ) &=& \int_{\partial AdS} dQ 
 \eta_{NN'}D^{NC} (d/2+c, Q) \mathcal{A}_{A_1A_2A_3C} (\Delta_1, P_1; \Delta_2, P_2;\Delta_3 , P_3;  d/2+c, Q|c') \nonumber\\
&&\times D^{N' C'} (h-c',Q)\mathcal{A}_{A_4 A_5 C'} (\Delta_4,P_4;\Delta_5, P_5; d/2-c, Q)
\ees

\begin{center}
\tikzset{
particle/.style={decorate, draw=black,
    decoration={coil,aspect=0.08,segment length=3pt,amplitude=3pt}}}\begin{tikzpicture}[thick, node distance=1cm and 1.5cm]
\coordinate[label={[xshift=-3pt]left:$P_1,{A_1}, a_1~~$}] (e1);

\coordinate[below right=of e1,label={[xshift=3pt]right:$~X$}] (aux1);

\coordinate[above right=of aux1,label={[xshift=6pt]right:$~~P_2,{A_2}, a_2$}] (e2);

\coordinate[below=1cm of aux1] (aux);

\coordinate[below left=of aux2,label={[xshift=-6pt]left:$P_5, {A_5}, a_5~~$}] (e3);

\coordinate[below right=of aux2,label={[xshift=3pt]right:$~~P_4,{A_4}, a_4$}] (e4);

\coordinate[below=1cm of aux1,label={[xshift=3pt]above left:$Z~~~$}] (aux3);

\coordinate[below=2cm of aux1,label={[xshift=3pt]right:$~Y$}] (aux2);

\draw[particle] (e1) -- (aux1);
\draw[particle] (aux1) -- (e2);
\draw[particle] (e3) -- (aux2);
\draw[particle] (aux2) -- (e4);
\draw[particle] (aux1) -- node {} (aux2);
\node[draw,name path=circle,line width=3pt,circle,fit=(e1) (e4),inner sep=.5\pgflinewidth] {};
\path[name path=diameter] let \p1=(aux1), \p2=(aux2) 
 in (aux1|-0,0.5*\y2+0.5*\y1) -- ++(3cm,0);
\path[name intersections={of=circle and diameter, by={aux3}}];

\path[name path=diameter] let \p1=(aux1), \p2=(aux2) 
 in (aux1|-0,0.5*\y2+0.5*\y1) -- ++(-3cm,0);
\path[name intersections={of=circle and diameter, by={aux4}}];

\draw[particle2] (aux) -- (aux4);
\draw[particle2] (aux2) -- (aux4);

\draw[particle] (aux) -- (aux3);
\node[label={[xshift=2pt]right:$P_3,{A_3},a_3$}] at (aux3) {};
\node[label={[xshift=2pt]left:$\int dQ~~ Q$}] at (aux4) {};

\end{tikzpicture}

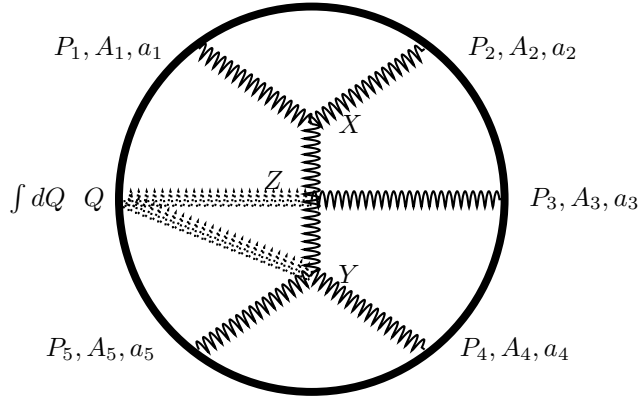
\captionof{figure}{The five-point vector amplitude \eqref{fivepoint}.}
\label{5ptvb}
\end{center}
The three-point amplitude in \eqref{eq118} is given by \eqref{eq48xa}.
The four-point amplitude simplifies to
\be D^{NC} (d/2+c, Q) \mathcal{A}_{A_1A_2A_3C} =\int (\mathcal{D}_4 \mathcal{M})^{N}_{A_1A_2A_3}  (\delta_{ij}' | c')  \prod_{i<j}  P_{ij}^{-\delta_{ij}'}  \Gamma (\delta_{ij}') d\delta_{ij}' \ee
where
\be (\mathcal{D}_4 \mathcal{M})^{N}_{A_1A_2A_3} = \frac{\frac{d}{2}+c -1}{\frac{d}{2}+c}\eta^{NC} \mathcal{M}_{A_1 A_2 A_3C}
- \frac{1}{\frac{d}{2}+c}  \left(   \delta_{14}'P_{1}^{N}\frac{P_{3}\cdot Q}{P_{1}\cdot Q} + \delta_{24}'P_{2}^{N}\frac{P_{3}\cdot Q}{P_{2}\cdot Q} + (\delta_{34}' -1) P_{3}^{N}\right) \frac{Q^{C}}{P_{3}\cdot Q} \mathcal{M}_{A_1 A_2 A_3C}   \ee
\bes \mathcal{M}_{A_1 A_2 A_3C} &=& \frac{(\frac{d}{2}-1)^2 -{c'}^2}{\frac{d^2}{4}-{c'}^2} \left[ \mathcal{I}({1,2,3,4})
\delta_{13}' + \mathcal{I} (1,2,4,3) \delta_{14}' - (1\longleftrightarrow 2)  \right]
\nextl
\times \frac{\prod_{\sigma=\pm}\Gamma (\frac{\frac{3d}{2}+\sigma c' -1}{2}) \Gamma ( \frac{d + c + \sigma c' }{2} ) \Gamma ( \delta_{12}' -  \frac{\frac{3d}{2}+\sigma c' -1}{2})}{\Gamma (\delta_{12}') \Gamma (\delta_{12}' +1-\frac{\frac{d}{2}  +h-c+1}{2})} \ees
\be \mathcal{I}({1,2,3,4}) = -\frac{ c-c'  }{d-2(c'+1)} \eta_{A_1A_2}\eta_{A_3C}
- 2 \eta_{A_1A_2} \frac{P_{1A_3} P_{3C}}{P_{13}} +4
\eta_{A_1A_3} \frac{P_{1A_2} P_{3C}}{P_{13}}
-2\frac{c-c' }{\frac{d}{2}-c'-1} \eta_{A_3C} 
\frac{P_{1A_2} P_{3A_1}}{P_{13}}
\ee
\bes \mathcal{I}({1, 2, 4,3}) &=&  \frac{ c+c'- d +2 }{d-2(c'+1)} \eta_{A_1A_2}\eta_{A_3C}
-\frac{2}{\delta_{34}' -1}\left( 2
\eta_{A_1A_4} P_{1A_2} - \eta_{A_1A_2} P_{1C} \right) \left( \delta_{13}' \frac{ P_{1A_3}}{P_{13}} +  \delta_{23}' \frac{ P_{2A_3}}{P_{23}} \right)\frac{P_{3}\cdot Q}{P_{1}\cdot Q}\nextl
-\frac{2}{\delta_{14}'}\frac{c+c' }{\frac{d}{2}-c'-1} \eta_{A_3C} 
\left( \frac{1}{2}\eta_{A_1A_2} + \delta_{12}' \frac{P_{2A_1}P_{1A_2}}{P_{12}} + \delta_{13}' \frac{P_{3A_1}P_{1A_2}}{P_{13}} \right)
\ees 
\bes\label{eq88a2} \frac{P_{13}}{-2P_{3}\cdot Q}Q^{C}\mathcal{I}({1, 2, 3,4}) &=&  \frac{ c-c'  }{(\delta_{34}' -1)(d-2(c'+1))} \eta_{A_1A_2}  \left( \delta_{13}' P_{1A_3} +  \delta_{23}' \frac{P_{13}}{P_{23}} P_{2A_3}  \right)
+ \eta_{A_1A_2} P_{1A_3}
+4
\eta_{A_1A_3} P_{1A_2} \nextl
+\frac{2}{\delta_{34}' -1}\frac{c-c' }{\frac{d}{2}-c'-1}
P_{1A_2}\left(  \frac{1}{2}\eta_{A_1A_3} + (\delta_{13}' +1)\frac{P_{3A_1}P_{1A_3}}{P_{13}} + \delta_{23}' \frac{P_{3A_1}P_{2A_3}}{P_{23}}  \right)\nextl
\ees
and
\bes\label{eq89a2}
\frac{P_{13}}{-2P_{3}\cdot Q}Q^{C}\mathcal{I}({1, 2, 4,3}) &=& -\frac{ c+c'- d +2 }{(\delta_{34}' -1)(d-2(c'+1))} \eta_{A_1A_2} \left(\delta_{13}'  P_{1A_3} +  \delta_{23}' \frac{P_{13}}{P_{23}} P_{2A_3}  \right) \nextl
-\frac{1}{\delta_{34}' -1} \eta_{A_1A_2} \left( \delta_{13}' P_{1A_3} +  \delta_{23}' \frac{ P_{13}}{P_{23}} P_{2A_3} \right)\nextl
+\frac{4\delta_{13}'}{\delta_{14}'(\delta_{34}' -1)}
   \left[  \frac{1}{2}\eta_{A_1A_2} P_{1A_3}+\frac{1}{2}\eta_{A_1A_3} P_{1A_2}+ \left( \delta_{12}' \frac{P_{2A_1}}{P_{12}} + (\delta_{13}' +1) \frac{P_{3A_1}}{P_{13}} \right) P_{1A_2}P_{1A_3} \right] \nextl
+\frac{4  \delta_{23}'}{\delta_{14}'(\delta_{34}' -1)}
 \frac{ P_{2A_3}}{P_{23}}P_{13}\left( \frac{1}{2}\eta_{A_1A_2} + \delta_{12}' \frac{P_{2A_1}P_{1A_2}}{P_{12}} + \delta_{13}' \frac{P_{3A_1}P_{1A_2}}{P_{13}} \right)\nextl
-\frac{2}{\delta_{14}'(\delta_{34}' -1)}\frac{c-c' }{\frac{d}{2}-c'-1}
\left( \eta_{A_1A_2} + \delta_{12} \frac{P_{2A_1}P_{1A_2}}{P_{12}}  \right) \left(\delta_{13}'  P_{1A_3} +  \delta_{23}' \frac{P_{13}}{P_{23}} P_{2A_3}  \right)\nextl
-\frac{2\delta_{13}'}{\delta_{14}'(\delta_{34}' -1)}\frac{c-c' }{\frac{d}{2}-c'-1}
  P_{1A_2} \left(  \frac{1}{2}\eta_{A_1A_3} + (\delta_{13}' +1)\frac{P_{3A_1}P_{1A_3}}{P_{13}} + \delta_{23}' \frac{P_{3A_1}P_{2A_3}}{P_{23}}  \right)\nextl
\ees 
The integral over $Q$ is of the same form as before (Eq.\ \eqref{eq101} with exponents given by \eqref{eq102}). However, this case is slightly more complicated because $n_i \in \{ 0, \pm 1, \pm 2\}$, $\sum_i n_i =0$ ($i=1,2,3$).
The integration variables are constrained by \eqref{eq245n}, and are given
in terms of the Mandelstam invariants by \eqref{eq104}.

\section{Higher-point amplitudes}
\label{secV}

In this section, we suppress group theory indices, as they are not involved in the calculations except as constant factors. Thus, but amplitude we mean a sub-amplitude with a given group theory structure.

Higher point amplitudes can be calculated recursively by sewing together diagrams. Consider two scalar diagrams with $N_1$ and $N_2$ external legs, respectively. Suppose they have been calculated and put in the form
\be A_{N_1s} (\Delta_1, P_1; \dots ; \Delta_{N_1} , P_{N_1} ) = \frac{g^{N_1-2}}{\prod_i 2\pi^{d/2} \Gamma (\Delta_i +1 - \frac{d}{2})} \int [dc'] \mathcal{A}_{N_1} ( \{ \Delta_i , P_i \} | [c'] )  \ee
with
\be \mathcal{A}_{N_1} ( \{ \Delta_i , P_i \} | [c'] ) = \frac{\pi^{d/2}}{2} \int \mathcal{M}_{N_1} (\delta_{ij}' | [c']) \prod_{i<j} \Gamma (\delta_{ij}') P_{ij}^{-\delta_{ij}'} d\delta_{ij}'  \ \ , \ \ \ \ \sum_{i\ne j} \delta_{ij}' = \Delta_i \ee
and similarly
\be A_{N_2s} (\Delta_{1}', P_{1}'; \dots ; \Delta_{N_2}' , P_{N_2}' ) = \frac{g^{N_2-2}}{\prod_i 2\pi^{d/2} \Gamma (\Delta_i' +1 - \frac{d}{2})} \int [dc''] \mathcal{A}_{N_2} ( \{ \Delta_i' , P_i' \} | [c''] )  \ee
with
\be \mathcal{A}_{N_2} ( \{ \Delta_i' , P_i' \} | [c''] ) = \frac{\pi^{d/2}}{2} \int \mathcal{M}_{N_2} (\delta_{ij}'' | [c'']) \prod_{i<j} \Gamma (\delta_{ij}'') (P_{ij}')^{-\delta_{ij}''} d\delta_{ij}''   \ \ , \ \ \ \ \sum_{i\ne j} \delta_{ij}'' = \Delta_i' \ee
After sewing together the last two legs in the respective diagrams, we create a $N$-point diagram with $N=N_1+N_2-2$. Its amplitude is given by
\be A_{Ns} (\Delta_1, P_1; \dots ; \Delta_{N} , P_{N} ) = \frac{g^{N-2}}{\prod_i 2\pi^{d/2} \Gamma (\Delta_i +1 - \frac{d}{2})} \int [dc'][dc'']\frac{dc}{2\pi i} f_{\delta ,0} (c) \mathcal{A}_N ( \{ \Delta_i , P_i \} | [c'],[c''],c )  \ee
with
\be \mathcal{A}_N = \int_{\partial AdS} dQ \mathcal{A}_{N_1} (\Delta_1, P_1; \dots ; \Delta_{N_1-1}, P_{N_1-1}; d/2+c , Q | [c'])
\mathcal{A}_{N_2} (\Delta_{N_1}, P_{N_1} ; \dots ; \Delta_N , P_N ; d/2-c, Q | [c'']) \ee
The integral over $Q$ involves
\be \int_{\partial AdS} dQ \prod_{i=1}^N \Gamma (\lambda_i) (-2Q\cdot P_i)^{-\lambda_i} = \frac{\pi^{d/2}}{2} \int \prod_{i<j} d\tilde\delta_{ij} \Gamma (\tilde\delta_{ij}) P_{ij}^{-\tilde\delta_{ij}} \ee
with $\sum_{j\ne i} \tilde\delta_{ij} = \lambda_i$, and
\be \lambda_i = \delta_{iN_1}' \ \ \ \ (i=1,\dots, N_1-1) \ \ , \ \ \ \ \lambda_{N_1+i} = \delta_{iN_2}'' \ \ \ \ (i=1,\dots, N_2-1) \ee
Then the part of the amplitude involving the vectors $P_i$ is
\be \left( \prod_{i<j}^{N_1-1} P_{ij}^{-\delta_{ij}'} \right) \left( \prod_{i<j}^{N_2-1} P_{N_1+i-1 \, N_1+j-1}^{-\delta_{ij}''} \right) \left( \prod_{i<j}^N P_{ij}^{-\tilde\delta_{ij}} \right) = \prod_{i<j} P_{ij}^{-\delta_{ij}} \ee
where $\delta_{ij}$ are the Mandelstam invariants for the $N$-point amplitude given by
\be \delta_{ij} = \tilde\delta_{ij} + \delta_{ij}' \ \ (i,j = 1,\dots, N_1-1) \ \ , \ \ \ \ \delta_{N_1+i-1 \, N_1+j-1} = \tilde\delta_{N_1+i-1 \, N_1+j-1} + \delta_{ij}'' \ \ (i,j = 1,\dots, N_2-1)~,\ee
and $\delta_{ij} = \tilde\delta_{ij}$, otherwise. They obey the constraints $\sum_{i\ne j} \delta_{ij} = \Delta_i$, as can easily be checked.

It follows that the $N$-point amplitude can be cast in the form
\be\label{eq126} \mathcal{A}_{N} = \frac{\pi^{d/2}}{2} \int \mathcal{M}_{N} \prod_{i<j} \Gamma (\delta_{ij}) P_{ij}^{-\delta_{ij}} d\delta_{ij}  \ee
where
\be \mathcal{M}_N = \frac{\pi^{d/2}}{2} \int d\delta_{ij}' \int d\delta_{ij}'' \prod_{i<j}^{N_1-1} \frac{\Gamma (\delta_{ij} - \delta_{ij}')\Gamma (\delta_{ij}')}{\Gamma (\delta_{ij})} \prod_{i<j}^{N_2-1} \frac{\Gamma (\delta_{N_1+i-1 \, N_1+j-1} - \delta_{ij}'')\Gamma (\delta_{ij}'')}{\Gamma (\delta_{N_1+i-1 \, N_1+j-1})} \mathcal{M}_{N_1} \mathcal{M}_{N_2}
\ee
The above procedure can be applied to the the case of vector amplitudes which can thus be written in the form \eqref{eq126}. In the vector case, a $N_1$-point diagram is given by
\be A_{N_1v}^{M_1\cdots M_{N_1}} (\Delta_1, P_1; \dots ; \Delta_{N_1} , P_{N_1} ) = \int [dc'] \prod_{i=1}^{N_1} D^{M_iA_i} (\Delta_i ,P_i) \mathcal{A}_{A_1\cdots A_{N_1}} ( \{ \Delta_i , P_i \} | [c'] )  \ee
with
\be \mathcal{A}_{A_1\cdots A_{N_1}} ( \{ \Delta_i , P_i \} | [c'] ) = \frac{\pi^{d/2}}{2} \int \mathcal{M}_{A_1\cdots A_{N_1}} (\delta_{ij}' | [c']) \prod_{i<j} \Gamma (\delta_{ij}') P_{ij}^{-\delta_{ij}'} d\delta_{ij}'  \ee
Similarly, a $N_2$-point diagram is given by
\be A_{N_2v}^{M_1\cdots M_{N_2}} (\Delta_1', P_1'; \dots ; \Delta_{N_1}' , P_{N_1}' ) = \int [dc''] \prod_{i=1}^{N_2} D^{M_iA_i} (\Delta_i' ,P_i') \mathcal{A}_{A_1\cdots A_{N_1}} ( \{ \Delta_i' , P_i' \} | [c'] )  \ee
with
\be \mathcal{A}_{A_1\cdots A_{N_1}} ( \{ \Delta_i' , P_i' \} | [c''] ) = \frac{\pi^{d/2}}{2} \int \mathcal{M}_{A_1\cdots A_{N_1}} (\delta_{ij}'' | [c'']) \prod_{i<j} \Gamma (\delta_{ij}'') (P_{ij}')^{-\delta_{ij}''} d\delta_{ij}''  \ee
By sewing together these two diagrams, we obtain a $N$-point vector diagram of amplitude
\be A_{Nv}^{M_1\cdots M_{N}} (\Delta_1, P_1; \dots ; \Delta_{N} , P_{N} ) = \int [dc'][dc'']\frac{dc}{2\pi i} f_{\delta ,1} (c) \prod_{i=1}^N D^{M_iA_i} (\Delta_i ,P_i) \mathcal{A}_{A_1\cdots A_{N}} ( \{ \Delta_i , P_i \} | [c'],[c''],c )  \ee
with
\bes\label{eq133} \mathcal{A}_{A_1\cdots A_{N}} &=& \int_{\partial AdS} dQ \eta_{MM'} D^{MC} (d/2+c, Q) \mathcal{A}_{A_1\cdots A_{N_1-1}C} (\Delta_1, P_1; \dots ; \Delta_{N_1-1}, P_{N_1-1}; d/2+c , Q | [c'])\nextl
\times D^{M'C'} (d/2-c,Q) \mathcal{A}_{A_1\cdots A_{N_2-1}C'} (\Delta_{N_1}, P_{N_1} ; \dots ; \Delta_N , P_N ; d/2-c, Q | [c'']) \ees
The integration over $Q$ can be performed in the same way as in the scalar case provided $Q$ only appears in dot products (as in $P_i\cdot Q$, and no terms with $Q$ with a free index exist). This can be ensured by the repeated application of the identity \eqref{eq42i4}, as we have already demonstrated. This leads to expressions for the two factors in \eqref{eq133}, $D^{MC}\mathcal{A}_{A_1\cdots A_{N_1-1}C}$ and $D^{M'C'} \mathcal{A}_{A_1\cdots A_{N_2-1}C'}$, respectively, containing no $Q$ with a free index.

After integrating over $Q$, we arrive at an expression for the amplitude of the form
\be\label{eq126v} \mathcal{A}_{A_1\cdots A_{N}} = \frac{\pi^{d/2}}{2} \int \mathcal{M}_{A_1\cdots A_{N}} \prod_{i<j} \Gamma (\delta_{ij}) P_{ij}^{-\delta_{ij}} d\delta_{ij}  \ee
where $\mathcal{M}_{A_1\cdots A_{N}}$ is given in terms of the same functions as in the scalar case.

To complete the iteration, we need to apply the identity \eqref{eq42i4} again, as many times as needed, on $D^{M_iA_i} \mathcal{A}_{A_1\cdots A_{N}}$, in order to eliminate all occurrences of $Q$ with a free index. The resulting expressions can then be used for the calculation of higher-point amplitudes.

\section{Conclusion}
\label{secVI}

We discussed an iterative method of calculation of Witten diagrams in AdS space based on the formalism developed in \cite{Paulos:2011ie}. We applied our method to scalar and vector fields and showed that they can both be written in terms of Mellin amplitudes which can be computed explicitly. We showed how this is done in detail for three-, four-, and five-point diagrams. We demonstrated that the index structure in the vector case did not present additional difficulties in the calculation of integrals over AdS space, by taking advantage of the conformal structure of the amplitudes.

Our method can be straightforwardly generalized to higher-spin fields (calculation of correlators of stress-energy tensors, etc). As it provides a systematic way of calculating diagrams, which appears to be uniformly applicable to fields of any spin, it would be interesting to use our method toward the development of general Feynman rules for the calculation of Witten diagrams.
Work in this direction is in progress.

\acknowledgments
We thank Miguel Paulos for all his helpful comments. Research supported in part by the Department of Energy under Grant No.\ DE-FG05-91ER40627.

\appendix
\section{Useful integrals}
\label{Xintegral}
Here we derive integrals over AdS space (bulk or boundary) that are used in our discussion.

Integrals on the boundary are often of the form
\be I = \int_{\partial AdS} dQ \prod_i \Gamma (\lambda_i) (-2P_i\cdot Q)^{-\lambda_i} \ee
where $Q$, $P_i$ are all points on the boundary and the exponents $\lambda_i$ are constrained
\be \sum_i \lambda_i = d \ee
To perform the integral over the vector $Q^A$, use the Mellin transform,
\be \Gamma (\lambda_i)  (-2P_i\cdot Q)^{-\lambda_i} = \int_0^\infty \frac{dv_i}{v_i} v_i^{\lambda_i} e^{2v_iP_i\cdot Q}\ee
to write the integral $I$ in the form
\be I = \int_0^\infty \prod_i \frac{dv_i}{v_i} v_i^{\lambda_i} \int_{\partial AdS} dQ e^{2T\cdot Q} \ \ , \ \ \ \ T^A = \sum_i v_i P_i^A \ee
Going to the rest frame of $T_A$ in which $T_A =  ( T_0 , 0,\dots, 0)$ with $T_0\ge 0$, and parametrizing the null vector $Q^A$ by
\be
Q^A= \left( \frac{x^2+1}{2},  \frac{x^2-1}{2}, x^\mu \right)
\ee 
we obtain
\be \int_{\partial AdS} dQ e^{2T\cdot Q} = \frac{\pi^{d/2}}{T_0^{d/2}} e^{-T_0} \ee
After rescaling $v_i \to v_i T_0$, we obtain
\be I = \pi^{d/2}\int_0^\infty \prod_i \frac{dv_i}{v_i} v_i^{\lambda_i} e^{T^2} \ \ , \ \ \ \ T^2 = - \sum_{i<j} v_iv_j P_{ij} ~, \ee
where we used $T^2 = - T_0^2$.

This expression can be simplified further by using the inverse Mellin transform
\be e^{-y} = \int_{c-i\infty}^{c+i\infty} \frac{ds}{2\pi i} \Gamma (s) y^{-s} \ \ \ \ (c>0) \ee
We obtain
\be I = \pi^{d/2}\int_0^\infty \prod_i \frac{dv_i}{v_i} v_i^{\lambda_i} \int_{c-i\infty}^{c+i\infty} \prod_{i<j} \frac{ds_{ij}}{2\pi i} \Gamma (s_{ij}) (v_iv_jP_{ij})^{-s_{ij}} ~.\ee
Each integral over $v_i$ yields a $\delta$-function. We deduce
\be\label{eqA10} I = \frac{\pi^{d/2}}{2} \int \prod_{i<j} \frac{ds_{ij}}{2\pi i} \Gamma (s_{ij}) P_{ij}^{-s_{ij}} ~,\ee
where the integration variables are constrained by
\be\label{eqA11} \sum_{j} s_{ij} = \lambda_i \ee
and we have defined $s_{ji} = s_{ij}$, $s_{ii} = 0$.

Integrals in the bulk are of a similar form,
\be\label{eqA12} J = \int_{AdS} dX \prod_i \Gamma (\Delta_i) (-2P_i\cdot X)^{-\lambda_i} = \int_0^\infty \prod_i \frac{dv_i}{v_i} v_i^{\lambda_i} \int_{ AdS} dX e^{2T\cdot X} \ee
where this time the exponents $\lambda_i$ are not constrained.

Parametrizing the bulk point $X^A$ by
\be
X= \frac{1}{x_0} \left( \frac{x_0^2 + x^2+1}{2},  \frac{x_0^2 + x^2-1}{2}, x^\mu \right)
\ee 
we obtain
\be \int_{AdS} dX e^{2T\cdot X} = \pi^{d/2} \int_0^\infty \frac{dx_0}{x_0} x_0^{-d/2} e^{-x_0+T^2/x_0} \ee
After rescaling $v_i \to v_i\sqrt{ x_0}$, the integral over $x_0$ can be performed, and we obtain
\be J = \pi^{d/2} \Gamma \left( \frac{\sum_i\lambda_i -d}{2} \right) \int_0^\infty \prod_i \frac{dv_i}{v_i} v_i^{\lambda_i} e^{T^2}  ~. \ee
The remaining integrals are simplified, as before,
\be\label{eqA16} J = \frac{\pi^{d/2}}{2} \Gamma \left( \frac{\sum_i\lambda_i -d}{2} \right)\int \prod_{i<j} \frac{ds_{ij}}{2\pi i} \Gamma (s_{ij}) P_{ij}^{-s_{ij}} ~,\ee
where the integration variables are constrained by \eqref{eqA11}.


\begin{thebibliography}{99}
\bibitem{Maldacena:1997re} 
  J.~M.~Maldacena,
  ``The Large N limit of superconformal field theories and supergravity,''
  Adv.\ Theor.\ Math.\ Phys.\  {\bf 2}, 231 (1998)
  [hep-th/9711200].
\bibitem{Witten:1998qj} 
  E.~Witten,
  ``Anti-de Sitter space and holography,''
  Adv.\ Theor.\ Math.\ Phys.\  {\bf 2}, 253 (1998)
  [hep-th/9802150].
\bibitem{Penedones:2010ue} 
  J.~Penedones,
  ``Writing CFT correlation functions as AdS scattering amplitudes,''
  JHEP {\bf 1103}, 025 (2011)
  [arXiv:1011.1485 [hep-th]].
\bibitem{Freedman:1998bj} 
  D.~Z.~Freedman, S.~D.~Mathur, A.~Matusis and L.~Rastelli,
  ``Comments on 4 point functions in the CFT / AdS correspondence,''
  Phys.\ Lett.\ B {\bf 452}, 61 (1999)
  [hep-th/9808006].

\bibitem{Liu:1998ty} 
  H.~Liu and A.~A.~Tseytlin,
  Phys.\ Rev.\ D {\bf 59}, 086002 (1999)
  [hep-th/9807097].
\bibitem{Freedman:1998tz} 
  D.~Z.~Freedman, S.~D.~Mathur, A.~Matusis and L.~Rastelli,
  ``Correlation functions in the CFT(d) / AdS(d+1) correspondence,''
  Nucl.\ Phys.\ B {\bf 546}, 96 (1999)
  [hep-th/9804058].
\bibitem{D'Hoker:1999ni} 
  E.~D'Hoker, D.~Z.~Freedman and L.~Rastelli,
  ``AdS / CFT four point functions: How to succeed at z integrals without really trying,''
  Nucl.\ Phys.\ B {\bf 562}, 395 (1999)
  [hep-th/9905049].
\bibitem{D'Hoker:1999pj} 
  E.~D'Hoker, D.~Z.~Freedman, S.~D.~Mathur, A.~Matusis and L.~Rastelli,
  ``Graviton exchange and complete four point functions in the AdS / CFT correspondence,''
  Nucl.\ Phys.\ B {\bf 562}, 353 (1999)
  [hep-th/9903196].
\bibitem{Raju:2010by}
  S.~Raju,
  ``BCFW for Witten Diagrams,''
Phys.\ Rev.\ Lett.\ {\bf 106}, 091601 (2011)
 [arXiv:1011.0780 [hep-th]].
\bibitem{Raju:2012zs}
  S.~Raju,
  ``Four Point Functions of the Stress Tensor and Conserved Currents in AdS4/CFT3,''
Phy.\ Rev.\ D {\bf 85}, 126008 (2012)
  [arXiv:1201.6452 [hep-th]].
\bibitem{Raju:2012zr}
  S.~Raju,
  ``New Recursion Relations and a Flat Space Limit for AdS/CFT Correlators,''
Phy.\ Rev.\ D {\bf 85}, 126009 (2012)
  [arXiv:1201.6449 [hep-th]].
\bibitem{Raju:2011mp}
  S.~Raju,
  ``Recursion Relations for AdS/CFT Correlators,''
  Phys.\ Rev.\ D {\bf 83}, 126002 (2011)
  [arXiv:1102.4724 [hep-th]].
\bibitem{Mack:2009mi} 
  G.~Mack,
  ``D-independent representation of Conformal Field Theories in D dimensions via transformation to auxiliary Dual Resonance Models. Scalar amplitudes,''
  arXiv:0907.2407 [hep-th].
\bibitem{Mack:2009gy} 
  G.~Mack,
  ``D-dimensional Conformal Field Theories with anomalous dimensions as Dual Resonance Models,''
  Bulg.\ J.\ Phys.\  {\bf 36}, 214 (2009)
  [arXiv:0909.1024 [hep-th]].
\bibitem{Fitzpatrick:2011ia} 
  A.~L.~Fitzpatrick, J.~Kaplan, J.~Penedones, S.~Raju and B.~C.~van Rees,
  ``A Natural Language for AdS/CFT Correlators,''
  JHEP {\bf 1111}, 095 (2011)
  [arXiv:1107.1499 [hep-th]].
\bibitem{Paulos:2011ie}
  M.~F.~Paulos,
  ``Towards Feynman rules
 for Mellin amplitudes,''
  JHEP {\bf 1110}, 074 (2011)
  [arXiv:1107.1504 [hep-th]].


\bibitem{Dirac:1936fq} 
  P.~A.~M.~Dirac,
  Annals Math.\  {\bf 37}, 429 (1936).
\bibitem{Weinberg:2010fx} 
  S.~Weinberg,
  Phys.\ Rev.\ D {\bf 82}, 045031 (2010)
  [arXiv:1006.3480 [hep-th]].
  \end{thebibliography}
\end{document}